\documentclass[12pt, a4paper] {article}
\usepackage{setspace}
\usepackage{amsmath}
\usepackage{amsfonts,epsfig}
\usepackage{amsthm}
\newtheorem{theorem}{Theorem}
\newtheorem{lemma}{Lemma}
\usepackage{amssymb}

\newtheorem{example}{Example}
\newtheorem{remark}{Remark}

\usepackage{multirow}
\usepackage{texnames}
\usepackage{ulem}
\usepackage{setspace}
\usepackage{paralist}
\usepackage{float}
\usepackage{bm}
\usepackage{color}

\makeatletter
\newcommand{\rmnum}[1]{\romannumeral #1}
\newcommand{\Rmnum}[1]{\expandafter\@slowromancap\romannumeral #1@}
\newcommand*{\rom}[1]{\expandafter\@slowromancap\romannumeral #1@}
\newcommand\red[1]{{}}
\newcommand\reda[1]{{}}
\makeatother

\usepackage{times}
\usepackage{bm}
\usepackage{natbib}

\usepackage{algorithm}
\usepackage{algorithmic}

\begin{document}
\thispagestyle{empty}

\title{Model-Free Sure Screening via Maximum Correlation}

\author{Qiming Huang and Yu Zhu\footnote{\baselineskip=10pt Qiming Huang is Ph.D candidate (e-mail: hqm@purdue.edu) and Yu Zhu is Professor
(e-mail: yuzhu@purdue.edu), Department of Statistics, Purdue University, West Lafayette, Indiana 47907.  The authors would like to thank all authors of  Li, Zhong and Zhu (2012) for the clarification of their technical proofs and all authors of Fan, Feng and Song (2011) for providing the R codes used in their paper.}}  

\date{}
\maketitle

\addtolength{\voffset}{-4cm}

\begin{abstract}
For screening features in an ultrahigh-dimensional setting, {we develop a maximum correlation-based sure independence screening (MC-SIS) procedure}, and show that MC-SIS possesses the sure screen property without imposing model or distributional assumptions on the response and predictor variables. MC-SIS is a model-free method as in contrast with some other existing model-based sure independence screening methods in the literature. Simulation examples and a real data application are used to demonstrate the performance of MC-SIS as well as to compare MC-SIS with other existing sure screening methods. The results show that MC-SIS can outperform those methods when their model assumptions are violated, and  remain competitive when the model assumptions are satisfied.


\end{abstract}

\addtolength{\voffset}{4cm}

\clearpage\pagebreak\newpage

\section{Introduction}
With rapid development of modern technology, various types of high-dimensional data are collected in a variety of areas such as next-generation sequencing and biomedical imaging data in bioinformatics, high-frequency time series data in quantitative finance, and spatial-temporal data in environmental studies.  In those types of high-dimensional data, the number of variables \textit{p} can be much larger than the sample size \textit{n}, which is referred to the `large $p$ small $n$' scenario.  To deal with this scenario, a commonly adopted approach is to impose the sparsity assumption that the number of important variables is small relative to $p$. Based on the sparsity assumption, a variety of regularization procedures have been proposed for high-dimensional regression analysis such as the lasso \citep{Tib96Lasso}, the smoothly clipped absolute deviation method \citep{FanLi01SCAD}, and elastic net \citep{ZouHastie05enet}.  All these methods work when $p$ is moderate.  However, when applied to analyze ultrahigh-dimensional data where dimensionality grows exponentially with sample size (e.g., $p = \exp(n^{\alpha})$ with $\alpha>0$), their performances will deteriorate in terms of computational expediency, statistical accuracy and algorithmic stability \citep{fan09ultra}.  To address the challenges of ultrahigh dimensionality, a number of marginal screening procedures have been proposed under different model assumptions.  They all share the same goal that is to reduce dimensionality from ultrahigh to high while retaining all truly important variables.  When a screening procedure achieves this goal, it is said to have the sure screening property in the literature.

\cite{FanLv08SIS} proposed to use Pearson correlation for feature screening and showed that the resulting procedure possesses the sure screening property under the linear model assumption. They refer to the procedure as the Sure Independence Screening (SIS) procedure. \cite{FanSong10GLM} extended SIS from linear models to generalized linear models by using maximum marginal likelihood values.  \cite{FanFengSong11NIS} developed a Nonparametric Independence Screening (NIS) procedure and proved that NIS has the sure screening property under the additive model. \cite{LiZhongZhu12Dcor} proposed to use distance correlation to rank the predictor variables, and showed that the resulting procedure, denoted as DC-SIS, has the sure screening property without imposing any specific model assumption. Compared with the other screening procedures discussed previously, DC-SIS is thus model-free. Distance correlation was introduced in \cite{Szekely07DC}, which uses joint and marginal characteristic functions to measure the dependence between two random variables.

From the review above, it is clear that the standard approach to developing a valid screening procedure consists of two steps. First, a proper dependence measure between the response and predictor variables needs to be defined and further used to rank-order all the predictor variables; and second, the sure screening property needs to be established for the screening procedure based on the dependence measure.  The screening methods discussed in the previous paragraph differ from each other in these two steps. For example, SIS uses Pearson correlation as the dependence measure and possesses the sure screening property under linear models, whereas NIS uses the goodness of fit measure of the nonparametric regression between the response and predictor variable as the dependence measure and possesses the sure screening property under additive models. 

For the purpose of screening in an ultrahigh dimensional setting, we argue that an effective screening procedure should employ a sensitive dependence measure and satisfy the sure screening requirement without model specifications.  The goal of screening is not to precisely select the true predictors, instead, it is to reduce the number of predictor variables from ultrahigh to high while retaining the true predictor variables. Therefore, false positives or selections can be tolerated to a large degree, and sensitive dependence measures are more preferred than insensitive measures. In ultrahigh dimensional data, there usually does not exist information about the relationship between the response and predictor variables, and it is extremely difficult to explore the possible relationship due to the presence of a large number of predictors.  Therefore, model assumptions should be avoided as much as possible in ultrahigh dimensional screening, and we should prefer screening procedures that possess the sure screening property without model specifications. In other words, model-free sure screening procedures are more preferable.  Among the existing screening procedures discussed previously, only DC-SIS does not require restrictive model assumption and therefore is model-free. However, the distance correlation measure used by DC-SIS may not be sensitive especially when sample size is small, because empirical characteristic functions are employed to estimate distance correlations. 

A more sensitive dependence measure between the response and a predictor variable is maximum correlation, which was originally proposed by \cite{Gebelein41} and studied by \cite{Renyi59MC} as a general dependence measure between two random variables.  \cite{Renyi59MC} gave a list of seven fundamental properties a reasonable dependence measure must have, and maximum correlation is one of a few measures that satisfy this requirement. The definition and estimation of maximum correlation involve maximizations over functions (see Section \ref{sec:mc_cal}), and thus it is fairly sensitive even when sample size is small. Recently, there have been some revived interests in using maximum correlation as a proper dependence measure in high-dimensional data analysis \citep{bickel09dc,hall11mc, reshef2011detecting,speed2011correlation}. 

In this paper, we propose to use maximum correlation as a dependence measure for ultrahigh dimensional screening, and prove that the resulting procedure has the sure screening property without imposing model specifications (see Theorem \ref{thm:ss} in Section \ref{sec:theory}). We adopt the B-spline functions-based estimation method \citep{Burman91ACE} to estimate maximum correlation.   We refer to our proposed procedure as the maximum correlation-based sure independence screening procedure, or in short, the MC-SIS procedure.  Numerical results show that MC-SIS is competitive to other existing model-based screening procedures, and is more sensitive and robust than DC-SIS when sample size is small or the distributions of the predictor variables have heavy tails. 

The rest of the paper is organized as follows. In Section \ref{sec:MC}, we introduce maximum correlation and the B-spline functions-based method for estimating maximum correlation, propose the MC-SIS procedure, and establish the sure screening property for MC-SIS.  In Section \ref{sec:tun}, we develop a three-step procedure for selecting tuning parameters for MC-SIS in practice.  Section \ref{sec:numeric} presents results from simulation study and a real life screening application.  Section \ref{sec:discussion} concludes the paper with additional remarks and future research. The proofs of the theorems are given in the Appendix.

\section{Independence Screening via Maximum Correlation}
\label{sec:MC}
\subsection{Maximum correlation and optimal transformation}
\label{sec:mc_cal}
Let $Y$ denote the response variable and $ \mathbf{X} = (X_1,\ldots, X_p)$ be the vector of predictor variables.  We assume the supports of $Y$ and $X_j$ $(j= 1,\ldots, p)$ are compact, and they are further assumed to be [0,1] without loss of generality.  For any given $j$, consider a pair of random variables ($X_j,Y$).  The maximum correlation coefficient between $X_j$ and $Y$, denoted as $\rho_j^*$, is defined as follows.
\begin{equation}
\rho_j^*(X_j,Y) =\underset{\theta, \phi}{{sup}}\lbrace \rho\left( \theta(Y),\phi(X_j)\right): 0 < {E} \{\theta^2(Y) \}< \infty, 0 < {E}\{ \phi^2(X_j) \}< \infty\rbrace,
\label{mc}
\end{equation} 
where $\rho$ is the Pearson correlation, and $\theta$ and $\phi$ are Borel-measurable functions of $Y$ and $X_j$.  We further denote $\theta_j^*$ and $\phi_j^*$ as the optimal transformations that attain the maximum correlation.

Maximum correlation coefficient enjoys the following properties \citep{Renyi59MC}: (a) $0 \leq \rho_j^*(X_j,Y) \leq 1$; (b) $\rho_j^*(X_j,Y) = 0$ if and only if $X_j$ and $Y$ are independent;   (c) $\rho_j^*(X_j,Y) = 1$ if there exist Borel-measurable functions $\theta^*$ and $\phi^*$ such that $\theta^*(Y) = \phi^*(X_j)$; and (d) if $X_j$ and $Y$ are jointly Gaussian, then $\rho_j^*(X_j,Y) = \mathopen|\rho(X_j,Y)\mathclose|$.  Some other properties of maximum correlation coefficient are discussed in  \cite{Szekely85MC}, \cite{Dembo01MC}, \cite{Bryc05MC}, and \cite{Yu08MC}.  Due to Property (d), it is clear that maximum correlation is a natural extension of Pearson correlation.  Note that Pearson correlation does not possess Properties (b) and (c).  For property (c), there are cases that Pearson correlation coefficient can be as low as zero when $Y$ is functionally determined by $X_j$.  For example, if $Y = X_1^2$ where $X_1 \sim \mathcal{N}(0,1)$, the Pearson correlation between $Y$ and $X_1$ is zero, whereas the maximum correlation is one.  Therefore, maximum correlation is a more proper measure of the dependence between two random variables than Pearson correlation.

\cite{Renyi59MC} established the existence of maximum correlation under certain sufficient conditions, and a different set of sufficient conditions are given in \cite{Breiman85JASA}.  \cite{Breiman85JASA} also showed that optimal transformations $\theta^*_j$ and $\phi^*_j$ can be obtained via the following minimization problem.
\begin{equation}
\begin{aligned}
& \underset{\theta_j, \phi_j \in L_2(P)}{\text{min}}
& &  e_j^{2} = {E}[\{\theta_j(Y) - \phi_j(X_j)\}^2], \\
& \text{subject to}
& & {E} \{\theta_j(Y)\} = {E}\{\phi_j(X_j)\} = 0; \\
&&& {E} \{\theta^2_j(Y)\} = 1.
\end{aligned}
\label{eq:opt}
\end{equation}
Here, \textit{P} denotes the joint distribution of ($X_j$,$Y$) and $L_2(P)$ is the class of square integrable functions under the measure \textit{P}.  Let $e_j^{*2}$ be the minimum of $e_j^2$.  \cite{Breiman85JASA} derived two critical connections between $e_j^{*2}$, squared maximum correlation $\rho_j^{*2}$, and optimal transformation $\phi_j^*$, which we state as \textit{Fact 0} below.
\begin{subequations}
\begin{align} 
\mbox{\textit{Fact 0.}} \hspace{2.2in} e_j^{*2} = 1 & - \rho_j^{*2};\hspace{1.8in}  
\label{eq:relation}\\
{E} ( \phi_j^{*2} ) & = \rho_j^{*2}.
\label{eq:relation2}
\end{align} 
\end{subequations}

\textit{Fact 0}  suggests that the minimization problem (\ref{eq:opt}) is equivalent to the optimization problem (\ref{mc}). Furthermore, the squared maximum correlation coefficient is equal to the expectation of the squared optimal transformation $\phi^*_j$.  

Various algorithms have been proposed in the literature to compute maximum correlation, including Alternating Conditional Expectations (ACE) in \cite{Breiman85JASA}, B-spline approximation in \cite{Burman91ACE}, and polynomial approximation in \cite{bickel09dc} and \cite{hall11mc}.  Equation (\ref{eq:relation2}) indicates that maximum correlation coefficient $\rho_j^*$ can be calculated through the optimal transformation $\phi_j^*$.  In this paper, we apply Burman's approach to first estimate $\phi^*_j$, and then estimate $\rho_j^*$, which will be further used in screening.

\subsection{B-spline estimation of optimal transformations}
\label{subsec:ot}

Let $\mathcal{S}_n$ be the space of polynomial splines of degree $\ell \geq 1$ and 
$\lbrace B_{jm},  m = 1,\ldots, d_n \rbrace$ 
denote a normalized B-spline basis with 
$\mathopen| \mathopen| B_{jm} \mathopen| \mathopen|_{{sup}} \leq 1$, where
$\mathopen| \mathopen|\cdot \mathopen| \mathopen|_{{sup}}$  is the sup-norm.  We have  
$\theta_{nj}(Y) = {\boldsymbol\alpha}_j^T \mathbf{B}_{j}(Y) $, $\phi_{nj}(X_j) = {\boldsymbol\beta}_j^T \mathbf{B}_{j}(X_j) $ for any $\theta_{nj}(Y), \phi_{nj}(X_j) \in \mathcal{S}_n$, where $\mathbf{B}_j(\cdot) = (B_{j1}(\cdot),\ldots , B_{jd_n}(\cdot))^T$ denotes the vector of $d_n$ basis functions.  Additionally, we let $k$ be the number of knots where $k = d_n - \ell$. The population version of B-spline approximation to the minimization problem (\ref{eq:opt}) can be written as follows. 
\begin{equation}
\begin{aligned}
& \underset{\theta_{nj}, \phi_{nj} \in \mathcal{S}_n}{\text{min}}
& &  {E}[\{ \theta_{nj}(Y) - \phi_{nj}(X_j)\}^2], \\
& \text{subject to}
& & {E} \{ \theta_{nj}(Y) \} = {E} \{ \phi_{nj}(X_j) \} = 0; \\
&&& {E} \{\theta^2_{nj}(Y)\} = 1.
\end{aligned}
\label{eq:opt_bs}
\end{equation}

\cite{Burman91ACE} applied a technique to remove the first constraint ${E} \{ \theta_{nj}(Y) \} = \\ {E} \{ \phi_{nj}(X_j) \} = 0$  in the  optimization problem above as follows.   First, let $\mathbf{z}_1,\ldots, \mathbf{z}_{d_n - 1} $ ($\mathbf{z}_i = (z_{i1}, \ldots, z_{id_n})^T$ for $i = 1, \ldots, d_n -1$) be $d_n$-dimensional vectors which are orthogonal to each other, orthogonal to the vector of 1's and $\mathbf{z}_i^{T} \mathbf{z}_i = 1$ for $i = 1,\ldots, d_n-1$.    Second, obtain matrix $\mathbf{D}_j$ with the $(s,m)$-entry $\mathbf{D}_{j,sm} = z_{sm}/(kb_{jm})$ where $b_{jm} = {E} \{ B_{jm}(X_{j}) \}$, for $s=1,\ldots, d_n-1$ and $m = 1,\ldots, d_n$.  Third, let $\phi_{nj}(X_j) =  {\boldsymbol\eta}_j^T \boldsymbol{\psi}_j(X_j)$ where $\boldsymbol{\psi}_j(X_j) = \mathbf{D}_j  \mathbf{B}_j(X_j) $.  With this construction, it is easy to verify that ${E} \{ \phi_{nj}(X_j) \} = 0$, and the minimization of $ {E}[\{ \theta_{nj}(Y) - \phi_{nj}(X_j) \}^2]$ subject to ${E} \{\theta^2_{nj}(Y)\} = 1$ ensures that ${E} \{ \theta_{nj}(Y) \} = 0$.   \cite{Burman91ACE} showed the equivalence between the optimization problem (\ref{eq:opt_bs}) and the one stated below.
\begin{equation}
\begin{aligned}
& \underset{\theta_{nj}, \phi_{nj} \in \mathcal{S}_n}{\text{min}}
& &  {E}[\{ \theta_{nj}(Y) - \phi_{nj}(X_j)\}^2], \\
& \text{subject to}
&& {E} \{\theta^2_{nj}(Y)\} = 1.
\end{aligned}
\label{eq:opt_bs0}
\end{equation}
For fixed $\theta_{nj}(Y)$ (i.e. fixed ${\boldsymbol\alpha}_j$), the minimizer of (\ref{eq:opt_bs0}) with respect to ${\boldsymbol\eta}_j$ and  $\phi_{nj}(X_j)$ are
\begin{equation}
\begin{aligned}
& {\boldsymbol\eta}_j = [ {E} \{ {\boldsymbol\psi}_j(X_j) {\boldsymbol\psi}_{j}^T(X_j) \} ] ^{-1} {E} \{ {\boldsymbol\psi}_j(X_j)  \mathbf{B}_{j}^T(Y) \} {\boldsymbol\alpha}_j,  \\
& \phi_{nj}(X_j) = {\boldsymbol\psi}_j^T(X_j) [ {E} \{ {\boldsymbol\psi}_j(X_j) {\boldsymbol\psi}_{j}^T(X_j) \} ] ^{-1} {E} \{ {\boldsymbol\psi}_j(X_j)  \mathbf{B}_{j}^T(Y) \} {\boldsymbol\alpha}_j. 
\end{aligned}
\label{eq:phi}
\end{equation}
By plugging $\phi_{nj}(X_j)$  back in (\ref{eq:opt_bs0}), we obtain the following maximization problem,
\begin{equation}
\begin{aligned}
& \underset{{\boldsymbol\alpha}_j \in \mathbb{R}^{d_n}}{\text{max}}
& & {\boldsymbol\alpha}_j^T {E} \{ \mathbf{B}_{j} (Y) {\boldsymbol\psi}_j^T(X_j) \} [ {E} \{ {\boldsymbol\psi}_j(X_j) {\boldsymbol\psi}_{j}^T(X_j)\}]^{-1} {E} \{ {\boldsymbol\psi}_j(X_j) \mathbf{B}_{j}^T(Y) \} {\boldsymbol\alpha}_j, \\
& \text{subject to}
& & {\boldsymbol\alpha}_j^T  {E} \{ \mathbf{B}_{j}(Y) \mathbf{B}_{j}^T(Y) \}  {\boldsymbol\alpha}_j = 1.
\end{aligned}
\label{eq:opt_bs2}
\end{equation}
Following the notation in \cite{Burman91ACE}, we denote 
\begin{equation*}
\begin{aligned}
\mathbf{A}_{j00} = {E} \{ \mathbf{B}_{j}(Y) \mathbf{B}_{j}^T(Y) \}, \qquad
\mathbf{A}_{jXX} = {E} \{ \boldsymbol{\psi}_{j}(X_j) \boldsymbol{\psi}_{j}^T(X_j) \}, \\
\mathbf{A}_{jX0} = {E}  \{ \boldsymbol{\psi}_{j}(X_j) \mathbf{B}_{j}^T(Y) \}, \quad \mbox{and} \quad \mathbf{A}_{j0X} = \mathbf{A}_{jX0}^T.
\end{aligned}
\end{equation*}
It is clear that (\ref{eq:opt_bs2}) is a generalized eigenvalue problem,  which can be solved by the largest eigenvalue and its corresponding eigenvector of $\mathbf{A}_{j00}^{-1/2} \mathbf{A}_{j0X} \mathbf{A}_{jXX}^{-1}  \mathbf{A}_{jX0} \mathbf{A}_{j00}^{-1/2}$.  We denote the largest eigenvalue by $\lambda_{j1}^*$, which is equal to $|| \mathbf{A}_{j00}^{-1/2} \mathbf{A}_{j0X} \mathbf{A}_{jXX}^{-1}  \mathbf{A}_{jX0} \mathbf{A}_{j00}^{-1/2}||$, where $||\cdot||$ is the operator norm, and further denote the corresponding eigenvector by ${\boldsymbol\alpha}^*_j$.  Let $\phi^*_{nj}(X_j) =   {\boldsymbol\psi}_j^T(X_j) [ {E} \{ {\boldsymbol\psi}_j(X_j) {\boldsymbol\psi}_{j}^T(X_j) \} ] ^{-1} {E} \{ {\boldsymbol\psi}_j(X_j)  \mathbf{B}_{j}^T(Y) \} {\boldsymbol\alpha}^*_j$.  $\phi^*_{nj}$ can be considered the spline approximation to the optimal transformation $\phi^*_j$ defined previously.  Note that the target function in (\ref{eq:opt_bs2}) is ${E}(\phi_{nj}^{*2})$, and we also have $ {E}(\phi_{nj}^{*2}) = \lambda_{j1}^*$.

Given the data $\{Y_u\}_{u=1}^n$ and $\{X_{uj}\}_{u=1}^n$, we estimate $\mathbf{A}_{j00}$, $\mathbf{A}_{jXX}$, $\mathbf{A}_{jX0}$, and $\mathbf{A}_{j0X}$ as follows. 
\begin{equation*}
\begin{aligned}
\widehat{\mathbf{A}_{j00}} = n^{-1} \sum_{u=1}^{n} \mathbf{B}_j(Y_u)\mathbf{B}_j^T(Y_u), \qquad
\widehat{\mathbf{A}_{jXX}} = n^{-1} \sum_{u=1}^{n} \widehat{\boldsymbol\psi}_j(X_{uj}) \widehat{\boldsymbol\psi}_j^T(X_{uj}), \\
\widehat{\mathbf{A}_{jX0}} = n^{-1} \sum_{u=1}^{n} \widehat{\boldsymbol\psi}_j(X_{uj}) \mathbf{B}_j^T(Y_u), \quad \mbox{and} \quad \widehat{\mathbf{A}_{j0X}} = \widehat{\mathbf{A}_{jX0}}^T,
\end{aligned}
\end{equation*}
where $\widehat{\boldsymbol\psi}_j(X_{uj}) = \widehat{\mathbf{D}_j} \mathbf{B}_j(X_{uj})$, the ($s,m$)-entry of $\widehat{\mathbf{D}_j}$ is $\widehat{\mathbf{D}}_{j,sm} = z_{sm}/(k \widehat{b_{jm}})$, and 
$\widehat{b_{jm}} = n^{-1} \sum_{u=1}^{n} B_{jm}(X_{uj})$, for $s = 1,\ldots, d_n - 1$ and  $m = 1,\ldots, d_n$.  Then, $\lambda_{j1}^*$ is estimated by
\begin{center}
$\widehat{\lambda_{j1}^*} = || \widehat{\mathbf{A}_{j00}}^{-1/2} \widehat{\mathbf{A}_{j0X}} \widehat{\mathbf{A}_{jXX}}^{-1}  \widehat{\mathbf{A}_{j0X}}^T \widehat{\mathbf{A}_{j00}}^{-1/2}||$,
\end{center} 
and ${\boldsymbol\alpha}^*_j$ is estimated by the eigenvector of $\widehat{\mathbf{A}_{j00}}^{-1/2} \widehat{\mathbf{A}_{j0X}} \widehat{\mathbf{A}_{jXX}}^{-1}  \widehat{\mathbf{A}_{j0X}}^T \widehat{\mathbf{A}_{j00}}^{-1/2}$ corresponding to $\widehat{\lambda_{j1}^*}$, which we denote as $\widehat{{\boldsymbol\alpha}^*_j}$.  Therefore,  the optimal transformation of $Y$ is estimated by $\widehat{\theta^*_{nj}} = \widehat{{\boldsymbol\alpha}^*_j}^T B_j(Y)$.  Furthermore, based on (\ref{eq:phi}), the optimal transformation of $X_j$ can be obtained  by $\widehat{\phi^*_{nj}} = \widehat{{\boldsymbol\eta}^*_j}^T {\boldsymbol{\psi}}_j(X_j)$ with $\widehat{{\boldsymbol\eta}^*_j} = \widehat{\mathbf{A}_{jXX}}^{-1}  \widehat{\mathbf{A}_{jX0}}  \widehat{{\boldsymbol\alpha}^*_j} $.

Based on the two relationships  ($\rmnum{1}$) ${E} ( \phi_j^{*2} )  = ( \rho_j^* )^2$ and ($\rmnum{2}$)  $  {E}(\phi_{nj}^{*2}) = \lambda_{j1}^*$, and the fact that $\phi_{nj}^*$ is the optimal spline approximation to $\phi_{j}^*$, we propose to screen important variables using the magnitudes of $\widehat{\lambda_{j1}^*}$ for $1\leq j \leq p$.

\subsection{MC-SIS procedure}
\label{sec:mc-sis}
Let $\nu_n$ be a pre-specified threshold, and $\widehat{\mathcal{D}_{\nu_n}} $ the collection of selected important variables.  Then our proposed screening procedure can be defined as
\begin{equation}
\widehat{\mathcal{D}_{\nu_n}} = 
\lbrace  1 \leq j \leq p  \colon  
\widehat{\lambda_{j1}^*} \geq \nu_n \rbrace.
\label{eq:select}
\end{equation}
\reda{Empirically, the threshold value $\nu_n$ is often set so that  $|\widehat{\mathcal{D}_{\nu_n}}| = n$  or  $\left[ n/\log n  \right]$, where $|\widehat{\mathcal{D}_{\nu_n}}|$ is the cardinality of $\widehat{\mathcal{D}_{\nu_n}}$ and} $\left[ a \right] $ denotes the integer part of $a$.  Since $\widehat{\lambda_{j1}^*}$ is the estimate of $\lambda_{j1}^*$, which is an approximation to the squared maximum correlation coefficient $\rho_j^{*2}$, we refer to the procedure as the MC-SIS procedure. The sure screening property of the procedure will be discussed in next section.

\subsection{Sure Screening Property}
\label{sec:theory}

Adopting notations from \cite{LiZhongZhu12Dcor}, we use $F(Y|\mathbf{X})$ to denote the conditional distribution of $Y$ given $\mathbf{X}$ and $\Psi_Y$ the support for $Y$. We define $\mathcal{D} = \lbrace j : F(y|\mathbf{X})$ functionally depends on $X_j$$\}$,  and $\mathcal{I} = \lbrace j : F(y|\mathbf{X})$ does not functionally depends on $X_j$$\}$. Let $\mathbf{X}_{\mathcal{D}} = \lbrace X_j$ : $j \in \mathcal{D} \rbrace$ and  $\mathbf{X}_{\mathcal{I}} = \lbrace X_j$ : $j \in \mathcal{I} \rbrace$, which are referred to the \textit{active}  and \textit{inactive sets}, respectively. Furthermore, we refer the variables in the active set and inactive set  as \textit{active predictor variables} and \textit{inactive predictor variables}, respectively.   Ideally, the goal of a screening procedure is to retain $\mathcal{D}$ after screening, which is referred to as the sure screening property.  We have established the sure screening property of the MC-SIS procedure under certain conditions. Before stating the theorem, we first list the conditions below.


\vspace{0.1in}
(C1) If the transformations $\theta_j$ and $\phi_j$ with zero means and finite variances satisfy
\begin{equation*}
 \theta_j(Y) + \phi_j(X_j) = 0  \mbox{ a.s., then each of them is zero a.s.}
\end{equation*}

(C2) The conditional expectation operators $ {E}\{\phi_j(X_j) \mid Y \} : H_2(X_j) \rightarrow H_2(Y)$  and ${E}\{ \theta_j(Y) \mid X_j \} : H_2(Y) \rightarrow H_2(X_j)$
are all compact operators.  $H_2(Y)$ and $H_2(X_j)$ are Hilbert spaces of all measurable functions with zero mean, finite variance and usual inner product.

(C3) The optimal transformations $\lbrace\theta^*_j, \phi^*_j \rbrace_{j = 1}^p$ belong to a class of functions $\mathcal{F}$, whose $\textit{r}$th derivative $\mathit{f}^{(r)}$ exists and is Lipschitz of order $\alpha_1$, that is, $\mathcal{F} = \lbrace \mathit{f} : |\mathit{f}^{(r)}(s) - \mathit{f}^{(r)}(t)| \leq K|s-t|^{\alpha_1} \mbox{ for all } s,t  \rbrace \mbox{ for some positive constant \textit{K}}$, 
where $\textit{r}$ is a nonnegative integer and $\alpha_1 \in (0,1]$ such that $d = \textit{r} + \alpha_1 > 0.5$.  

(C4) The joint density of $Y \mbox{ and } X_j$ $(j = 1, \ldots, p)$ is bounded and the marginal densities of $Y$ and $X_j$ are bounded away from zero.

\reda{(C5) The number of B-spline basis functions $d_n$ satisfies that $d_n \leq \mathop{\min}\limits_{j \in \mathcal{D}} (\rho_j^{*2}) / ( 2 c_1  n^{-2\kappa}),$  for some constant $c_1 > 0$  and constant $\kappa$ such that $0 \leq \kappa < d/(2d + 1)$.}

(C6) There exist positive constant $C_1$ and constant $\xi \in (0,1)$ such that $d_n^{-d-1} \leq c_1 (1-\xi) n^{-2\kappa}/C_1$.

\vspace{0.1in}

Conditions (C1) and (C2) are adopted from \cite{Breiman85JASA}, which ensure that the optimal transformations exist.  Conditions (C3) and (C4) are from \cite{Burman91ACE}, but modified for our two-variable scenario.   {Condition (C5) above is similar to Condition 3 in \cite{FanLv08SIS}, Condition C in \cite{FanFengSong11NIS}, and Condition (C2) in \cite{LiZhongZhu12Dcor}, which all require that the dependence between the response and active predictor variables cannot be too weak.   
We note that this condition is necessary, since a marginal screening procedure will fail when the marginal dependence between the response and an active predictor variable is too weak.  

The following lemma shows that the maximum correlations achieved by B-spline-based transformations are at the same level as the original maximum correlations. 

\begin{lemma}
\label{lemma1}
Under conditions (C3) -- (C6), we have $ \mathop{\min}\limits_{j \in \mathcal{D}} \lambda_{j1}^* \geq c_1 \xi d_n n^{-2\kappa}$. 
\end{lemma} 

Based on condition (C1) -- (C6), we establish the following sure screening properties for MC-SIS.

\begin{theorem}
(a) Under conditions (C1) -- (C4), for any $c_2 > 0$, there exist positive constants  $c_3$ and $c_4$ such that
\begin{equation}
P( \mathop{\max}\limits_{1\leq j \leq p} | \widehat{\lambda_{j1}^*} - \lambda_{j1}^* |  \geq c_2 d_n n^{-2 \kappa} ) \leq  \mathcal{O} \left( p \zeta(d_n, n) \right).   
\label{thm:parta}
\end{equation}
where $\zeta(d_n, n) =  d_n^2 \exp ( - c_3 n^{1-4\kappa} d_n^{-4} )  +  d_n \exp ( -c_4 n d_n^{-7}) $.

(b) Additionally, if conditions (C5) and (C6) hold, by taking $\nu_n = c_5 d_n n^{-\kappa}$ with $c_5 \leq c_1 \xi/2$, we have that 
\begin{equation}
P ( \mathcal{D} \subseteq \widehat{\mathcal{D}_{\nu_n}})  \geq  1- \mathcal{O} \left( s \zeta(d_n, n) 
\right),  
\label{thm:partb} 
\end{equation} 
where $s$ is the cardinality of $\mathcal{D}$.

\label{thm:ss} 
\end{theorem}

Note that Theorem \ref{thm:ss} is stated for fixed number of predictor variables $p$.  In fact, the same theorem holds for divergent number of predictor variables $p_n$.  As long as $p_n \zeta(d_n, n)$ goes to zero asymptotically, MC-SIS can possess the sure screening property.  And we remark that the number of basis functions $d_n$ affects the final performance of MC-SIS.  To obtain the sure screening property, an upper bound of $d_n$ is $o(n^{1/7})$.  Since $d_n$ is determined by the choices of the degree of B-spline basis functions and the number of knots, different combinations of degree and the number of knots can lead to different screening results.  Additionally, knots placement can further affect the behavior of B-spline functions, and in practice, knots are usually equally spaced or placed at sample quantiles.  In next section, we will propose a data-driven three-step procedure for determine $d_n$ for MC-SIS in practice.  The optimal choice of $d_n$ and knots placement are beyond the scope of this paper and can be an interesting topic for future research.}

The sure screening property from Theorem \ref{thm:ss} guarantees that MC-SIS retains the active set.  The size of the selected set can be much larger than the size of the active set.  Therefore, it is of interest to assess the size of the selected set, similar to \cite{FanFengSong11NIS}. The next theorem is such a result for MC-SIS.
\begin{theorem}
Under Conditions (C1) -- (C6), we have that for any $\nu_n = c_5 d_n n^{-\kappa}$, there exist positive constants  $c_3$ and $c_4$ such that 
\begin{equation}
P [ |\widehat{\mathcal{D}_{\nu_n}}| \leq \mathcal{O}\{ n^{2\kappa} \lambda_{max}(\mathbf{\Sigma}) \} ]  \geq  1- \mathcal{O} \left( p_n \zeta(d_n, n) 
\right),  
\end{equation} 
where $|\widehat{\mathcal{D}_{\nu_n}}|$ is the cardinality of $\widehat{\mathcal{D}_{\nu_n}}$, \reda{$\lambda_{max}(\mathbf{\Sigma})$ is the largest eigenvalue of $\mathbf{\Sigma}$}, $\mathbf{\Sigma}  = E \{ {\boldsymbol\psi} {\boldsymbol\psi}^T\} $, $ {\boldsymbol\psi} = ( {\boldsymbol\psi}^T_1, \ldots, {\boldsymbol\psi}^T_{p_n}  )^T$,   $p_n$ is the divergent number of predictor variables, and  $\zeta(d_n, n)$  is defined in Theorem \ref{thm:ss} .  
\label{thm:fs} 
\end{theorem} 

From Theorem \ref{thm:fs}, we have that when $\lambda_{max}(\mathbf{\Sigma}) = \mathcal{O}(n^{\tau})$, the cardinality of the selected set by MC-SIS will be of order  $\mathcal{O}(n^{2\kappa +\tau})$.  Thus, by applying MC-SIS, we can reduce dimensionality from the original exponential order to a polynomial order, while retaining the entire active set.

\section{Tuning Parameter Selection}
\label{sec:tun}
\reda{
In the previous section, we showed that to achieve the sure screening property of MC-SIS, we need to impose several conditions on the choice of $d_n$.  Recall $d_n= k+\ell$, where $k$ is the number of knots, and $\ell$ is the degree of the B-spline basis functions. These conditions are of theoretical interest, however, they cannot be directly implemented in practice. It is well known that the performance of B-spline functions in nonparametric regression depends on the choices of $k$ and $\ell$ as well as the placement of knots. This is also the case for the performance of MC-SIS under a given finite sample. 

Several rules of thumb have been proposed to choose $d_n$ for B-spline basis functions when used for the purpose of screening in the literature. For example, cubic splines with $d_n=\left[ n^{1/5} \right] + 2$ were used in \cite{FanFengSong11NIS},  and cubic splines with  $ d_n= \left[ 2n^{1/5} \right] $ were proposed in \cite{fan2013nonparametric}, and in both works, the knots were placed at the sample quantiles. These rules of thumb can also be applied to MC-SIS, however, we found their performances \ are not so satisfactory in some models we have experimented with. In this section, we propose a more effective approach
for selecting $\ell$ and $k$ (or $d_n$) of the B-spline basis functions used in MC-SIS.

There are two major factors in the use of B-spline basis functions, which affect the performance of MC-SIS. The first factor is the complexity of the B-spline basis functions characterized by $\ell$ and $k$. The larger $\ell$ and $k$ are, the more complex the B-spline basis functions. Using more complex basis functions can clearly lead to the overfitting problem for inactive predictor variables, many of which may be retained due to their falsely inflated empirical correlations with the response variable. On the other hand, using less complex basis functions with small $\ell$ and $k$ can lead to the underfitting problem for active predictor variables, that is, the maximum correlations between the response variable and active predictor variables may be underestimated, and some active predictor variables may be ranked lower due to underestimated maximum correlations. Therefore, the proper selection of $\ell$ and $k$ hinges on the balance between the overfitting and underfitting problems. 

The other factor that affects the performance of MC-SIS is whether the same choices of $\ell$ and $k$ are used for all predictor variables, which is referred to as the unified scheme, or different choices of $\ell$ and $k$ are used for different predictor variables, which is referred to as the separate scheme. The unified scheme treats all predictor variables the same way, and is relatively simple. However, the unified scheme may be appropriate for some variables but inappropriate for other variables.  It is difficult to find a unified scheme that is proper for all predictor variables. On the other hand, the separate scheme allows individual variables to choose their most suitable basis functions. The separate scheme has however two drawbacks. The first drawback is it may exacerbate the over-fitting problem for inactive predictor variables, and the second is its computing is intensive. 

Based on the discussion above, it is clear that 
for the purpose of screening, an ideal scheme for choosing basis functions for MC-SIS is to use the unified scheme with simple basis functions for inactive predictor variables and the separate scheme with complex basis functions for active predictor variables. This ideal scheme is not feasible in practice because we do not know which predictor variables are active and which inactive ahead of time. In what follows next, we instead propose a data-driven three-step approach to approximate the ideal scheme. Because B-splines of degree higher than 3 are seldom used in practice, we only consider $\ell \in \{1, 2, 3\}$. Furthermore, we always place knots at  sample quantiles. 

In the first step, we use the unified scheme with B-spline basis functions of degree 1. In other words, we fix $\ell=1$. The number of knots $k$ is then determined as follows. Consider a set of candidate values for $k$, for example, $K_1 \leq k \leq K_2$ 
, where $K_1$ and $K_2$ are 
pre-specified integers. For each $k$, we first calculate the maximum correlations between the response variable and the predictor variables using $k$ knots and $\ell=1$, and then we fit a two-component Gaussian mixture distribution to the maximum correlations and denote the resulting component means as $\mu_1(k)$ and $\mu_2(k)$, respectively. Let $d(k)=| \mu_1(k)-\mu_2(k)|$, and $\tilde{k}=\min_{K_1 \le k\le K_2} d(k)$. Then we apply MC-SIS with $\ell=1$ and $k=\tilde{k}$ and retain $B_1$ predictor variables, where $B_1$ is a pre-specified number. The purpose of using the unified scheme with linear B-spline basis functions in this step is to avoid the over-fitting problem and screening out the majority of inactive predictor variables. 

In the second step, we employ the separate scheme. For each remaining predictor variable,  a $M$-fold cross validation (CV) procedure  is used to select $\ell \in\{1, 2\}$ and 
$k$ (where $K_1 \leq k \leq K_2$), and then the maximum correlation between the predictor variable and the response variable is calculated using B-spline basis functions with the selected $\ell$ and $k$. Subsequently, we rank-order the predictors using their corresponding maximum correlations, and retain the top $B_2$ predictor variables. The $M$-fold CV procedure uses the correlation between the response variable and the predictor variable as the score function, and the procedure is standard in \cite{friedman2009elements}. The purpose of using the separate scheme and B-spline basis functions of higher degree is to correct the under-fitting problem possibly suffered by the active predictor variables in the first step.  

The third step is similar to the second step. The only difference is that the degree $\ell$ for B-spline basis functions is selected from $\{1, 2, 3\}$ instead of $\{1, 2\}$. In other words, for individual remaining predictor variables, B-spline basis functions of degree up to 3 may be used to calculate their maximum correlations. The purpose of using cubic spline basis functions is to provide sufficient capacity to calculate the maximum correlations of active predictor variables. Once the maximum correlations are calculated, the predictor variables are sorted and the top $B_3$ are retained as the final output of MC-SIS. 

}

\section{Numerical Results}
\label{sec:numeric}
We illustrate the MC-SIS procedure by studying its performance under different model settings and distributional assumptions of the predictor variables.  For all examples, we compare MC-SIS with SIS, NIS, and DC-SIS.  As mentioned at the end of Section \ref{sec:mc_cal}, the ACE algorithm in \cite{Breiman85JASA} can also be used to calculate maximum correlation coefficient. Therefore, the ACE algorithm can also be used to perform maximum correlation-based screening, and we refer to the resulting procedure as the ACE-based MC-SIS procedure.  We also include the ACE-based MC-SIS procedure in our simulation study. To avoid confusion, we refer to our proposed procedure as the B-spline-based MC-SIS procedure in this section.  For each simulation example, we set $p = 1000$ and choose  $n  \in \lbrace 200, 300, 400 \rbrace$.

{Following \cite{FanLv08SIS} and \cite{FanFengSong11NIS}, we measure the effectiveness of MC-SIS  using minimum model size (MMS) and robust estimate of its standard deviation (RSD).  MMS is defined as the minimum number of selected variables, i.e., the size of the selected set, that is required to include the entire active set.  RSD is defined as IQR/1.34, where IQR is the interquartile range.}  When constructing B-spline basis functions, \reda{we choose the  degree and the number of knots according to the procedure proposed  in Section \ref{sec:tun}, and set  $K_1 = 3$, $K_2 = 6$, $B_1= 200$, $B_2 = 50$ and $M=10$.}

\begin{example} 
\label{example1}
\textit{(1.a):} $Y = {\boldsymbol\beta}^{*T} \mathbf{X} + \varepsilon$, with the first \textit{s} components of $\boldsymbol\beta ^{\ast }$ taking values $\pm 1$ alternatively and the remaining being 0, where $s = 3, 6 \mbox{ or } 12$; $ X_k $ are independent and identically distributed as $N\left( 0,1\right)$ for $1 \leq k \leq 950$; $ X_k  = \sum _{j=1}^{s}X_{j}\left( -1\right) ^{j+1} / 5+ ( 1- s \varepsilon_{k}/25)^{1/2}$ where $\varepsilon_k$ are independent and identically distributed as $N\left( 0,1\right)$ for $k = 951, \ldots, 1000$;  and $\varepsilon \sim N\left( 0,3\right)$.  {Here, $\mathcal{D} = \{ 1,2, \ldots, s\}.$}

\textit{(1.b):}  $Y = X_1 + X_2 + X_3 + \varepsilon$, where $ X_k$ are independent and identically distributed as $N\left( 0,1\right)$ for $k=1, \mbox{and } 3 \leq k \leq 1000$; $X_2 = \dfrac{1}{3} X_1^3 + \tilde{\varepsilon}$, and $\tilde{\varepsilon} \sim N\left( 0,1\right)$; and $\varepsilon \sim N\left( 0,3\right)$.  {Here, $\mathcal{D} = \{ 1, 2, 3\}$.}
   
\end{example}

The first example is from \cite{FanFengSong11NIS} and the simulation results are presented in Table \ref{tb:ex1}.  Under model (1.a), SIS demonstrates the best performance across all cases, which is expected since SIS is specifically developed for linear models. Under the models (1.a) with $s = 3 \mbox{ or } 6$, when $n = 200$, MC-SIS under-performs all other methods.  However, when sample size increases to 300 or 400, MC-SIS becomes comparable to others.  For the case with $s = 12$, MC-SIS under-performs other methods for all choices of $n$.  The cause for the relatively poor performance of MC-SIS is due to the weak signal.  With $s=12$, it requires more samples for MC-SIS to estimate maximum correlation coefficient, without taking advantages of linearity assumptions.  

In model (1.b), SIS fails as there exists a nonlinear relationship between $X_1$ and $X_2$.  NIS demonstrates the best performance as NIS is designed for dealing with nonparametric additive models.  The ACE-based MC-SIS procedure demonstrates the second best performance.  The B-spline-based MC-SIS procedure performs  better than DC-SIS.
  
\begin{table}
\def~{\hphantom{0}}
\caption{ MMS and RSD (in parenthesis) for Example \ref{example1}}{
    \begin{tabular}{l|r|r|r|r|r|rcc}
\hline\hline
        Model  & n   & SIS & NIS & DC-SIS & MC-SIS  & MC-SIS \\
         &&&&& (ACE) & (B-spline)\\
\hline 
        1.a  & 200 & \textit{5.8(3.0)}& 6.4(3.0)  & 6.8(3.2) & 11.9(7.7) & 36.6(20.7) \\ 
        (s = 3) & 300 & \textit{4.6(0.9)} & 4.9(1.5)   & 5.1(1.5)  & 5.9(3.0) & 15.0(6.7)\\ 
          & 400 & \textit{3.3(0.0)} & 3.4(0.0) & 3.6(0.8) & 3.6(0.8) & 6.8(3.7)\\ 
\hline 
         1.a  & 200 & \textit{57.4(2.4)} & 68.7(9.7)& 60.2(3.7)& 140.5(60.8) & 175.0(50.2) \\ 
        (s = 6) & 300 & \textit{56.0(0.0)} & 58.2(0.2)  & 57.1(0.0)   & 67.4(5.2)   & 94.7(27.8)\\ 
          & 400 & \textit{55.8(0.0)} & 55.9(0.0) & 55.9(0.0)& 56.8(0.8) & 68.0(9.0)\\         
\hline 
       1.a  & 200 & \textit{119.4(42.9)}& 250.6(133.2)& 195.2(55.8) &484.6(181.9)& 500.4(197.4)\\ 
        (s = 12) & 300 & \textit{73.4(7.5)}&120.6(35.3)& 80.3(10.6)  & 211.2(108.4)& 248.9(103.9)\\ 
          & 400 & \textit{64.5(0.8)}& 82.21(6.7)&69.7(1.5) &118.2(90.8)& 178.2(41.2)\\ 
\hline 
        1.b  & 200   & 443.6(455.2) & \textit{26.5(6.7)} & 136.1(113.4) & 56.8(32.8) & 115.7(84.7) \\ 
        & 300   & 394.5(379.7)   & \textit{7.3(0.0)}&  59.9(48.5)     & 21.9(5.4) & 51.9(27.4) \\ 
         & 400   &  410.0(361.2)  & \textit{3.2(0.0)}  & 41.1(36.8)  & 5.6(0.8) & 20.0(4.7) \\
\hline 
    \end{tabular}}
  \label{tb:ex1}
  \end{table}

\begin{example}
\label{example2}\textit{(2.a):} $Y = X_1X_2 + X_3X_4 + \varepsilon$;  {$\mathcal{D} = \{1,2,3,4 \}$;}
 \textit{(2.b):} $Y = X_1^2 + X_2^3 + X_3^2X_4 + \varepsilon$;   {$\mathcal{D} = \{ 1,2,3,4\}$;}  \textit{(2.c):} $Y = X_1\sin(X_2) + X_2\sin(X_1) + \varepsilon$;  {$\mathcal{D} = \{1,2 \}$;} \textit{(2.d):}  $Y = X_1\exp(X_2) + \varepsilon$;  {$\mathcal{D} = \{ 1,2\}$;} \textit{(2.e):} $Y = X_1\log(|c_0 + X_2|) + \varepsilon$; {$\mathcal{D} = \{ 1,2\}$;}  \textit{(2.f):} $Y = X_1/(c_0 + X_2) + \varepsilon$; {$\mathcal{D} = \{ 1,2\}$.} Here $X_{1},\ldots ,X_{1000}$ and $\epsilon$ are generated independently from $N(0,1)$,  and $c_0=10^{-4}$. 
\end{example}

The eight models considered in this example are non-additive, and the simulation results are presented in Table \ref{tb:ex2}.  Due to the presence of non-additive structures, we notice that SIS and NIS fail in all models, and increasing sample size does not help improve the performances of  SIS and NIS for most models.   Both MC-SIS and DC-SIS work well in this example, but MC-SIS outperforms DC-SIS for almost all the models in terms of MMS.  Even when the sample size is as small as 200, MC-SIS can effectively retain the active set under models (2.c), (2.e) and (2.f).  This example demonstrates the advantages of MC-SIS and DC-SIS over SIS and NIS for non-additive models as well as the effectiveness of MC-SIS over DC-SIS. 

\begin{table}
\def~{\hphantom{0}}
\caption{MMS and RSD (in parenthesis) for Example \ref{example2}}{
\begin{tabular}{l|r|r|r|r|r|rr}
\hline\hline 
      Model   & n   & SIS & NIS & DC-SIS & MC-SIS  & MC-SIS \\
         &&&&& (ACE) & (B-spline)\\
\hline 
   2.a & 200 & 709.3(239.0)& 651.5(285.5)& 440.6(231.2) & \textit{248.7(242.5)} & 324.3(228.2)\\ 
         & 300 & 724.1(194.6) & 631.2(251.7) & 350.5(186.0) & \textit{117.8(88.3)}& 197.8(152.6)\\ 
        & 400 & 795.3(194.8)& 636.5(256.3)& 280.0(148.9) & \textit{59.3(26.1)} & 118.2(92.2)\\
\hline 
   2.b & 200 & 617.5(308.2)& 300.5(298.7)& 186.5(132.5)& \textit{104.2(103.0)} & 176.5(135.1) \\ 
         & 300 & 608.5(305.0) & 277.8(250.0)& 163.6(150.2) & \textit{78.4(44.6)}& 125.1(71.6) \\ 
         & 400 & 597.4(291.6) & 262.0(228.9) & 114.7(103.7)& \textit{54.9(13.9)}& {63.8(32.1)} \\         
\hline 
   2.c & 200 & 574.5(352.2)&511.7(389.0)& 113.6(80.2)& \textit{18.1(2.24)} &  {30.9(15.1)}\\ 
         & 300 & 616.4(342.2)&521.8(321.6)& 51.0(30.0) & \textit{8.4(0.8)} & {9.6(3.2)}\\ 
         & 400 & 622.4(306.3)& 547.8(337.9)&21.4(14.0) &13.0(0.0)& \textit{4.8(2.2)} \\ 
\hline 
   2.d & 200 &536.5(285.1)&181.8(168.5)& \textit{2.0(0.0)} & 2.3(0.8) & 9.7(3.2)\\ 
         & 300 & 268.6(307.1)& 172.8(190.9)& \textit{2.0(0.0)} & \textit{2.0(0.0)}& {6.4(3.0)} \\ 
          & 400 & 272.1(331.0) & 176.3(178.7) &\textit{2.0(0.0)} & \textit{2.0(0.0)}& {4.7(2.2)} \\
\hline 
   2.e  & 200   & 580.2(152.8) & 512.2(405.6) & 191.0(152.8) & 55.1(20.3) & \textit{26.6(14.2)} \\ 
          & 300   & 588.7(299.4)   & 641.0(295.3)& 107.1(70.3) & 40.7(1.5) & \textit{11.5(4.5)} \\ 
          & 400   &  602.1(258.4)  & 568.0(311.9) & 66.2(44.6)  & 19.8(0.0) & \textit{7.6(3.7)}\\ 
\hline 
   2.f  & 200   & 928.8(59.3) & 654.5(417.9) & 140.5(123.5) & \textit{30.0(9.9)} & {40.8(11.9)} \\ 
          & 300   & 936.7(37.7)   & 768.8(292.0)& 61.6(46.6) & {23.4(2.2)} & \textit{17.5(6.0)}\\ 
          & 400   &  942.0(39.9) & 821.7(175.2) & 60.9(22.8) & {17.8(0.8)} & \textit{12.6(3.7)} \\
\hline         
    \end{tabular}
    }
  \label{tb:ex2}
  \end{table}

\begin{example}  The models considered in this example are modifications of the models considered in Example \ref{example2}.  First, the error term $\epsilon$ in each original model is removed; and second, the predictor variables $X_1, X_2, \ldots, X_p$ are drawn independently from $Cauchy(0,1)$ instead of $N(0,1)$. The resulting models are denoted as (3.a)-(3.f), correspondingly. Simulation results based on these models are presented in Table \ref{tb:ex3}.
\label{example3}
\end{example}

Intuitively, the absence of the error terms in the models is expected to help the screening methods, but the use of heavy-tailed distributions for the predictor variables is expected to hinder the methods. The exact performance of a screening method in this example depends on the trade-off between those two changes. Comparing Table \ref{tb:ex3} with Table \ref{tb:ex2}, we can see that the performances of SIS and NIS have improved, though they are still far from being satisfactory. The performance of DC-SIS has improved in models (3.a) and (3.c), but has much deteriorated in the other models, which indicates that DC-SIS is susceptible to heavy-tailed distributions. In the presence of heavy tails, Condition (C1) in \cite{LiZhongZhu12Dcor} is violated, and DC-SIS may not have the sure screening property.  The performances of ACE-based and B-spline-based MC-SIS are better over DC-SIS in most models, which indicates the robustness of MC-SIS towards heavy-tailed distributions.

\begin{table}
\def~{\hphantom{0}}
\caption{MMS and RSD (in parenthesis) for Example \ref{example3}}{  
    \begin{tabular}{l|r|r|r|r|r|r} \hline\hline
       Model   & n   & SIS & NIS & DC-SIS & MC-SIS  & MC-SIS \\
         &&&&& (ACE) & (B-spline)\\
    \hline 
        3.a & 200 & 338.8(284.3)& 296.6(175.4)& {90.3(54.3)} & 124.1(39.2) &\textit{78.7(26.5)}\\ 
         & 300 & 310.2(241.6) & 310.8(253.7) & {64.6(32.5)} & 72.2(14.7) & \textit{44.6(9.1)}\\ 
        & 400 & 273.3(242.4)& 303.1(260.6)& 48.3(29.9) & \textit{41.5(7.1)} & \textit{34.5(6.0)}\\
    \hline 
         3.b & 200 & 617.5(305.2)& 617.5(256.7)& 478.9(286.6)& {117.8(36.6)} & \textit{79.6(56.0)}\\ 
         & 300 &  665.8(348.3)&689.2(256.2)& 511.2(258.8)& {72.0(8.6)} & \textit{42.1(6.2)}\\ 
         & 400 & 619.8(297.0) & 696.8(250.0) & 507.8(265.1)& {32.7(6.7)} & \textit{32.2(6.9)} \\         
    \hline 
         3.c & 200 & 136.5(80.2)&106.6(70.7)& 23.7(12.7)& \textit{11.9(5.2)} & {22.8(6.9)}\\ 
         & 300 & 116.1(82.1)& 90.1(56.2)& 13.4(6.3) & \textit{8.7(4.5)} & {17.3(6.2)}\\ 
         & 400 & 90.4(36.0)& 67.9(39.2)& 9.9(4.7) & \textit{7.3(3.2)} & {13.7(5.2)}\\ 
    \hline 
         3.d & 200 &409.5(367.0) & 434.8(409.0) & 412.3(401.1) & \textit{15.4(3.7)} & {19.3(6.0)}\\ 
         & 300 & 485.1(320.0) & 486.7(411.0) & 493.8(397.0) & \textit{7.8(2.4)} & {14.1(3.7)}\\ 
          & 400 & 460.8(342.0) & 493.4(360.1) &480.7(407.3)& {12.5(0.0)}  & \textit{11.5(3.7)}\\
    \hline 
        3.e  & 200   & 252.2(193.8) & 250.2(228.5) & 124.0(99.1) & {55.8(11.4)} & \textit{39.6(8.2)} \\ 
          & 300   & 332.9(332.7)   & 340.0(289.0)& 188.7(120.9) & {42.9(4.5)} & \textit{36.1(7.5)} \\ 
          & 400   &  314.3(315.5)  & 334.6(308.6) & 121.1(98.0)  & {37.8(4.1)}& \textit{22.8(6.0)}\\ 
    \hline 
        3.f  & 200   &779.8(172.0) & 737.0(244.2) & 507.7(249.6) & {37.5(6.9)} & \textit{27.4(6.0)} \\ 
          & 300   & 808.4(149.8)   & 855.7(120.9)& 498.6(336.0) & {28.7(4.5)} & \textit{20.7(5.2)}\\ 
          & 400  &806.7(149.1) & 837.6(143.5)& 432.6(281.9)& 34.3(3.7) & \textit{17.3(3.9)}\\
    \hline 
    \end{tabular} }
  \label{tb:ex3}
\end{table}
 
\begin{example}  In this example, we consider a real data set that contains the expression levels of 6319 genes and the expression levels of a G protein-coupled receptor (Ro1) in 30 mice \citep{segal2003regression}. The same data set has been analyzed in \cite{HallMiller09GenerCor} and in \cite{LiZhongZhu12Dcor} using DC-SIS. The goal is to identify the most influential genes for Ro1.  
\label{example4} 
\end{example}

{We apply SIS, NIS, DC-SIS, ACE-based MC-SIS and B-spline-based MC-SIS to select the top two most important genes, separately.   \reda{For B-spline-based MC-SIS, as the number of observations is small, we set $K_1 = 1$, $K_2 = 4$, $B_1 = 100$, $B_2 = 30$ and $M=3$ for the parameter selection procedure in Section \ref{sec:tun}.  B-spline-based MC-SIS ranks   \textit{Msa.2437.0} and \textit{Msa.26751.0} as the top two genes.  We note that gene \textit{Msa.2437.0} is ranked at the second place by ACE-based MC-SIS  and in the  15th place by NIS.  Gene \textit{Msa.26751.0} is ranked in the 22nd place by ACE-based MC-SIS and in the 41st place by SIS. } 
Additionally, we note that almost all of the procedures considered here, including B-spline-based MC-SIS, consistently ranked \textit{Msa.741.0}, \textit{Msa.2134.0} and \textit{Msa.2877.0} among the top ranked genes. The top-ranked two genes by individual procedures are reported in Table \ref{tb:ex4}.

To further compare the performances of the screening procedures, we fit regression models for the response, which is the expression level of Ro1, using the top two genes selected by the procedures. Three different models are considered, which are the linear regression model
$ Y = \beta_0 + \beta_{1} X_{1} + \beta_{2} X_{2} + \varepsilon $, 
the additive model 
$ Y = \ell_{1}(X_{1}) + \ell_{2}(X_{2}) + \varepsilon $, 
and the optimal transformation model 
$ \theta^*(Y) = \phi^*_{1}(X_{1}) + \phi_{2}^* (X_{2}) + \varepsilon$,
where $\theta^*$, $\phi_1^*$ and $\phi_2^*$ are the optimal transformations \citep{Breiman85JASA}.  For each procedure, all  three models are fitted using the top ranked gene as well as using the top ranked two genes, and the resulting adjusted $R^2$ values are reported in Table \ref{tb:ex4a}.
\begin{table}\begin{center}
\caption{Top ranked genes for Example \ref{example4}}{
  \begin{tabular}{c|c|c|c|c|c}
      \hline     \hline 
   & SIS & NIS & DC-SIS & MC-SIS & MC-SIS \\
   &&&& (ACE) & (B-spline)\\
    \hline 
   Rank 1 gene  & Msa.2877.0 & Msa.2877.0 & Msa.2134.0  & Msa.8081.0 & Msa.2437.0\\
    \hline 
   Rank 2 gene  &  Msa.964.0 & Msa.1160.0  & Msa.2877.0 & Msa.2437.0 & Msa.26751.0\\
  \hline 
  \end{tabular}
  }
  \label{tb:ex4}
\end{center}\end{table}

\begin{table}\begin{center}
\caption{ Adjusted $R^2$ (in percentage) of fitting 3 different models for Example \ref{example4}}{
\begin{tabular}{c|cc|cc|cc|cc|cc}    
\hline     \hline 
& \multicolumn{2}{ c| }{SIS} & \multicolumn{2}{ c| }{NIS} & \multicolumn{2}{ c |}{DC-SIS} & \multicolumn{2}{ c| }{MC-SIS} & \multicolumn{2}{ c}{MC-SIS} \\ 
&&&&&&&\multicolumn{2}{ c| }{(ACE)} & \multicolumn{2}{ c }{(B-spline)} \\

Model &  top 1 & top 2 &  top 1 & top 2 &  top 1 & top 2 &  top 1 & top 2 & top 1 & top 2 \\
    \hline 
Linear    & 74.5  & 82.3  & 74.5 &  75.8\ & 58.4 & 77.6  &  13.8  & 16.9  &  12.7  &  40.5 \\ 
    \hline 
Additive  & 80.0  & 84.2  & 80.0  & 84.5  & 65.7  & 96.8   &  58.9  & 68.7 &  68.5  & 68.8 \\
    \hline 
Transformation & 84.5  & 88.1  & 84.5  & 88.0  & 90.0  & 94.7   & 94.1  & 96.9  & 94.1  & 96.2\\ 
    \hline 
\end{tabular}}
  \label{tb:ex4a}
\end{center}\end{table}

Under the linear model, as expected, SIS achieves the largest adjusted $R^2$ values, whereas the adjusted $R^2$ values of ACE-based MC-SIS are rather poor. The major cause for the difference between SIS and ACE-based MC-SIS is that the former is specifically developed for screening under the linear model, whereas the latter is for screening under the optimal transformation model. Under the additive model, when the top one gene is used, NIS achieves the largest adjusted $R^2$ value; and when the top two genes are used, DC-SIS achieves the largest adjusted $R^2$ value. Under the optimal transformation model, MC-SIS (both ACE-based and B-spline-based) methods achieve the largest adjusted $R^2$ values with both the top one gene and top two genes.  When plotting the expression levels of Ro1 against the expression levels of various selected genes, different patterns including linear and nonlinear patterns emerge for different screening methods. In practice, we believe that the top ranked genes by different methods are all worth further investigation.

\section{Discussion}
\label{sec:discussion}
\qquad  The performances and results of B-spline-based MC-SIS depend on the choice of degree and the number of knots for B-splines.  In this paper, we have developed a data-driven three-step procedure to construct B-spline basis functions for MC-SIS in practice. The proposed procedure demonstrates satisfactory performance in simulation study as well as real data application.  We hope to investigate and characterize the theoretical property  of the procedure in the future.  

{Similar to other existing screening procedures, MC-SIS fail to retain active predictor variables that are marginally independent with the response variable.  Under the linear regression model,  \cite{FanLv08SIS} proposed an iterative procedure to recover such predictor variables.  Similarly, we have developed an iterative version of MC-SIS with the hope to recover active predictor variables missed by MC-SIS.  Currently, we are investigating the empirical performance and theoretical property of this iterative version and hope to report the results in a future publication.

Most existing marginal screening procedures under nonparametric model assumptions, including MC-SIS, make use of independent measures, whose estimation typically involves nonparametric model fitting and tuning parameter selection.  Nonparametric methods are known to be sensitive to tuning parameter selection.  Therefore, this can also become a drawback for those screening procedures.  On the other hand, there are various independence measures that are based on cumulative distribution functions, and the estimation of those measures does not involve nonparametric fitting and tuning parameter selection.  Two examples include Hoeffding's test \citep{hoeffding1948non} and Heller-Heller-Gorfine tests \citep{heller2012consistent}.  It will be of interest to explore the application of these measures for screening and the potential of using these methods for variable selection after screening.}

\pagebreak

\appendix
\begin{center}
      \Large{\bf Appendix}
    \end{center}
\section{Proofs}

\label{sec:proofs}
\subsection{Notation}

 \quad  $n$ : sample size

$p$ : dimension size

$\ell$ : degree of polynomial spline

$k$ : number of knots

$d_n$ : dimension of B-spline basis

$\mathcal{D}$ : active set

$\mathcal{I}$ : inactive set

$\theta_j$ : transformation of response $Y$ for pair $\left(X_j,Y \right)$, $j = 1,2,\ldots,p$

$\phi_j$ : transformation of $X_j$ for pair $\left(X_j,Y \right)$

$\rho_j$ : Pearson correlation of pair $\left(X_j,Y \right)$

$e_j^2$ : squared error by regressing $\phi_j$ on $\rho_j$

$\theta_j^*$ : optimal transformation of response $Y$ for pair $\left(X_j,Y \right)$

$\phi_j^*$ : transformation of $X_j$ for pair $\left(X_j,Y \right)$

$\rho_j^*$ : maximum correlation of pair $\left(X_j,Y \right)$

$e_j^{*2}$ : squared error  by regressing $\phi_j^*$ on $\theta_j^*$

$\theta_{nj}^*$ : spline approximation to optimal transformation $\theta_j^*$

$\phi_{nj}^*$ : spline approximation to optimal transformation $\phi_j^*$







$s$ : cardinality of active set $\mathcal{D}$

$\mathopen| \mathopen| \cdot \mathopen| \mathopen|$ : operator norm

$\mathopen| \mathopen| \cdot \mathopen| \mathopen|_{{sup}}$ : sup norm

\subsection{Bernstein's inequality and four facts}

\begin{lemma} (Bernstein's inequality, Lemma 2.2.9, \cite{van96weak})  For independent random variables $Y_1, \ldots, Y_n$ with bounded ranges $\left[-M,M\right]$ and 0 means,
\begin{equation*}
  P \left(  | Y_1 + \ldots + Y_n | > x \right) 
  \leq 2 \exp [ -x^2 / \{2(v + Mx/3) \} ] 
\end{equation*}
for $v \geq var( Y_1 + \ldots + Y_n)$.
\label{lemma2}
\end{lemma}

Under conditions (C3) and (C4), the following four facts hold when $\ell \geq d$.

\textit{Fact 1.} There exists a positive constant $C_1$ such that \citep{Burman91ACE}
\begin{equation}
    {E} \{ (\phi^*_{j} - \phi^*_{nj}  )^2 \} \leq C_1 k^{-d}
    \label{fact1}
\end{equation}

\textit{Fact 2.} There exists a positive constant $C_2$ such that \citep{stone85additive, huang10AOS}   
\begin{equation}
   {E} \{ B_{jm}^2(\cdot) \} \leq C_2 d_n^{-1}
\label{fact2}
\end{equation}
 
\textit{Fact 3.}  There exist positive constants $c_{11}$, $c_{12}$ such that \citep{Burman91ACE,zhou98AOS}   
\begin{equation}
\begin{aligned}
    c_{11} d_n^{-1} \leq \lambda_{min}\left( {E} \{ \mathbf{B}_j(\cdot) \mathbf{B}_j^T(\cdot) \}\right) &\leq  \lambda_{max}\left( {E} \{ \mathbf{B}_j(\cdot) \mathbf{B}_j^T(\cdot) \} \right) \leq c_{12} d_n^{-1} \\
     c_{11} k^{-1} \leq \lambda_{min}\left( {E} \{ {\boldsymbol\psi}_j(X_j) {\boldsymbol\psi}_j^T(X_j)\} \right)&\leq \lambda_{max}\left( {E} \{ {\boldsymbol\psi}_j(X_j) {\boldsymbol\psi}_j^T(X_j)\} \right) \leq c_{12} k^{-1} 
    \label{fact3}
\end{aligned}
\end{equation}

\textit{Fact 4.}  
There exists a positive constant $C_3 $ such that \citep{Burman91ACE,faouzi99}
 \begin{equation}
 \begin{aligned}
     C_3 k^{-1} \leq  b_{jm}  \leq 1, \quad
 \quad 0 \leq \widehat{b_{jm}} \leq 1
 \end{aligned}
 \label{fact4}
 \end{equation} 
 
\begin{remark} 
The choice of knots plays a role in establishing the sure screening property.  When the knots of the B-splines are placed at the sample quantiles, $\widehat{b_{jm}}$ is positive.  When knots are uniform placed, $\widehat{b_{jm}}$ can be zero with a small probability.  According to \citet*[section 6a]{Burman91ACE}, when the marginal density $f_{X_j}(x) > \gamma_0 > 0$ by condition (C4) for each $X_j$, we have $ P ( \widehat{b_{jm}}= 0 \mbox{ for some }  m = 1,\ldots, d_n ) \leq k \exp ( - \gamma_0 n/k )$.  The results in \cite{Burman91ACE} are based on equally spaced knots, and our proof for MC-SIS use the same choice of knots, as the probability of $\widehat{b_{jm}}$ being zero is a small probability, we just acknowledge $\widehat{b_{jm}} > 0$ in the proof.  In fact, sure screening property still hold when the event $\widehat{b_{jm}}= 0$ is included.
\end{remark}

\begin{remark}
 With $\ell$ fixed, $k$ and $d_n$ are of the same order, we replace $k$ with $d_n$ in the following proof for convenience.
\end{remark}

\subsection{Proof of Lemma 1}

\begin{proof}
By Cauchy-Schwarz inequality, we have
\begin{equation*}
 {E} (\phi^{*2}_{j}) \leq
   2 {E} \{(\phi^*_{j} - \phi^*_{nj} )^2\} + 2 {E} (\phi^{*2}_{nj})    
\end{equation*}

Therefore,
\begin{equation*}
  {E} (\phi^{*2}_{nj}) \geq   \frac{1}{2} {E}  (\phi^{*2}_{j})  - {E} \{ (\phi^*_{j} - \phi^*_{nj} )^2 \}  
\end{equation*}
Lemma 1 follows from condition (C5) together with $  {E} (\phi^{*2}_{nj}) = \lambda_{j1}^*$.
\end{proof}

\subsection{Proof of eight basic results}

We list and prove eight results ({R1}) -- ({R8}) which together form the major parts in proving sure screening property of MC-SIS. For the rest of the paper, we use ${P}_n$ to denote the sample average.

{R1}. \quad With $c_{11}$ in \textit{Fact 3}, we have that, 
\begin{equation}
||\mathbf{A}_{j00}^{-1/2}|| \leq c_{11}^{-1/2} d_n ^{1/2} 
\label{eq:R1}
\end{equation}
\begin{proof}
$||\mathbf{A}_{j00}^{-1/2}|| = \lambda_{min}^{-1/2}(\mathbf{A}_{j00})$, result follows by \textit{Fact 3}.
\end{proof}

{R2}. \quad There exist positive constant $c_{13}$ such that 
\begin{equation}
||\mathbf{A}_{j0X}|| \leq c_{13} d_n ^{-1/2}
\label{eq:R2}
\end{equation}
\begin{proof} Let $\mathbf{u} = (u_1, \ldots, u_{d_n})^T \in R^{d_n}$ with $\sum_{m=1}^{d_n} u_m^2 = 1$. 
\begin{equation*}
\begin{aligned}
 \mathbf{u}^T {E} \{ \mathbf{B}_j(X_j) \mathbf{B}_j^T(Y) \} & {E} \{\mathbf{B}_j(Y) \mathbf{B}_j^T(X_j)  \} \mathbf{u} = \sum\limits_{i=1}^{d_n} \left[ \int \lbrace \sum\limits_{m=1}^{d_n} u_m B_{jm}(X_j)\rbrace B_{ji}(Y)  dF \right]^2  \\
 &\leq  \int \{ \sum\limits_{m=1}^{d_n} u_m B_{jm}(X_j) \}^2 dF  \times \sum\limits_{i=1}^{d_n} \lbrace \int B_{ji}^2(Y)  dF \rbrace \\ 
 &\leq \lambda_{max} [
 {E} \{ \mathbf{B}_j(X_j) \mathbf{B}_j^T(X_j) \} ]  \times d_n  \max\limits_{i} {E} \{ B_{ji}^2(Y)\}
 \end{aligned}
\end{equation*}
 Then, $|| {E} \{ \mathbf{B}_{j}(Y) \mathbf{B}_j^T(X_j) \}|| \leq ( c_{12} C_2/ d_n) ^{1/2} $ by \textit{Fact 2} and \textit{Fact 3}.  
 
It can be easily shown that, for $\mathbf{u} \in R^{d_n-1}$ with $\sum_{i=1}^{d_n-1} u_i^2 = 1$,
\begin{equation*}
\begin{aligned}
\mathbf{u}^T \mathbf{D}_j \mathbf{D}_j^T \mathbf{u} = \sum\limits_{m=1}^{d_n} \frac{1}{k^2 b_{jm}^2} \left( \sum\limits_{i=1}^{d_n - 1} u_i z_{im} \right)^2 \leq C_3^{-2} \sum\limits_{m=1}^{d_n}  \left( \sum\limits_{i=1}^{d_n - 1} u_i z_{im} \right)^2 \leq  C_3^{-2}
\end{aligned}
\end{equation*}
which indicates $||\mathbf{D}_j^T|| \leq C_3^{-1}$.  

Then, $||\mathbf{A}_{j0X}|| \leq   || {E} \{ \mathbf{B}_{j}(Y) \mathbf{B}_j^T(X_j) \}|| \hspace{0.05in} ||\mathbf{D}_j^T||  \leq c_{13} d_n^{-1/2}$ with $c_{13}=(c_{12} C_2)^{1/2} C_{3}^{-1}$.
\end{proof}

{R3}. \quad  For any given constant $c_4$, there exists a positive constant $c_8$ such that 
\begin{equation}
 P\{ ||\widehat{\mathbf{A}_{j00}}^{-1/2}||  \geq \left( (c_8 +1)c_{11}^{-1} d_n \right)^{1/2} \}  \leq 2 d_n^2 \exp ( -c_4 n d_n^{-3} )
\label{eq:R3} 
\end{equation}
\begin{proof} 
\quad  Since $||\widehat{\mathbf{A}_{j00}}^{-1/2}|| = \sqrt{||[{P}_n \{ \mathbf{B}_j(Y) \mathbf{B}_j^T(Y)\}]^{-1}||}$.  {R3} can be obtained via equation (26) in \cite{FanFengSong11NIS}, which is  $P\{ ||[{P}_n \{ \mathbf{B}_j(Y) \mathbf{B}_j^T(Y)\}]^{-1}||  \geq (c_8 +1) c_{11}^{-1} d_n \}  \leq 2 d_n^2 \exp ( -c_4 n d_n^{-3} )$.
\end{proof}

{R4}. \quad  There exist some positive constants $c_6$, $c_{7}$ such that, 
\begin{equation}
 P\{ ||\widehat{\mathbf{A}_{j0X}} || \geq c_{6} d_n^{-1/2} \}  \leq 4 d_n^2 \exp ( -c_7 n d_n^{-2} )
\label{eq:R4}
\end{equation}

\begin{proof} 
\quad As $||\widehat{\mathbf{A}_{j0X}} || = || {P}_n \{ \mathbf{B}_j(Y) \mathbf{B}_j^T(X_j) \} \widehat{\mathbf{D}_j}^T || \leq ||{P}_n \{ \mathbf{B}_j(Y) \mathbf{B}_j^T(X_j)\}|| \hspace{0.05in} ||\widehat{\mathbf{D}_j}^T||$, we firstly deal with $||{P}_n \{ \mathbf{B}_j(Y) \mathbf{B}_j^T(X_j) \}||$. 

For any square matrix $\mathbf{A}$ and $\mathbf{B}$, $||\mathbf{A} + \mathbf{B}||  \leq ||\mathbf{A}|| +||\mathbf{B}||$.  We have 
\begin{equation*}
\begin{aligned}
||\mathbf{A}|| - ||\mathbf{B}|| \leq  ||\mathbf{A} - \mathbf{B}||\mbox{\qquad and \quad}  
||\mathbf{B}|| -||\mathbf{A}||\leq  ||\mathbf{B} - \mathbf{A}||  
\end{aligned}
\end{equation*}

Then, 
\begin{equation*}
\mid ||\mathbf{A}|| -||\mathbf{B}||\mid  \leq ||\mathbf{A} - \mathbf{B}|| 
\end{equation*}

Let $\mathbf{V}_j = {P}_n \{ \mathbf{B}_j(Y) \mathbf{B}_j^T(X_j) \} - {E} \{ \mathbf{B}_j(Y) \mathbf{B}_j^T(X_j)\}$.  It follows that, 
\begin{equation*}
\mid ||{P}_n \{ \mathbf{B}_j(Y) \mathbf{B}_j^T(X_j) \}|| -||{E}\{ \mathbf{B}_j(Y) \mathbf{B}_j^T(X_j) \}|| \mid \leq ||\mathbf{V}_j||  
\end{equation*}

It is easy to verify that, 
\begin{equation*}
\mid  ||{P}_n \{ \mathbf{B}_j(Y) \mathbf{B}_j^T(X_j) \}||  - || {E}\{ \mathbf{B}_j(Y) \mathbf{B}_j^T(X_j) \}||\mid  \leq d_n ||\mathbf{V}_j||_{sup}
\end{equation*}

Since $||B_{jm}(\cdot)||_{sup} \leq 1$ and using \textit{Fact 2}, we have 
\begin{equation*}
\mbox{var}( B_{jm_1}(Y) B_{jm_2}(X_j)) \leq {E}  \{ B_{jm_1}^2(Y) B_{jm_2}^2(X_j) \} \leq {E} \{ B_{jm_1}^2(Y)\} \leq C_2d_n^{-1}
\end{equation*}

 By Bernstein's inequality, for any $\delta > 0$, 
\begin{equation}
P\{| ({P}_n - {E}) \{ B_{jm_1}(Y) B_{jm_2}(X_j) \}| \geq \delta/n \} \leq 2 \exp \{- \frac{\delta^2}{2(C_2 n d_n^{-1} + 2\delta/3)}  \}
\label{eq:R4a}
\end{equation}

Therefore, 
\begin{equation*}
P\{\mid ||{P}_n \{ \mathbf{B}_j(Y) \mathbf{B}_j^T(X_j) \})|| - ||{E}\{ \mathbf{B}_j(Y) \mathbf{B}_j^T(X_j) \}||\mid \geq d_n \delta/n \} \leq 2 d_n^2 \exp \{- \frac{\delta^2}{2(C_2 n d_n^{-1} + 2\delta/3)}  \}
\end{equation*}

Recalling {R2}, we have,
\begin{equation*}
 P\{  || {P}_n \{ \mathbf{B}_j(Y) \mathbf{B}_j^T(X_j) \}||  \geq d_n \delta/n + (c_{12} C_2/d_n)^{1/2} \} \leq 2 d_n^2 \exp \{- \frac{\delta^2}{2(C_2 n d_n^{-1} + 2\delta/3)}  \}
\end{equation*}

By taking $\delta = c_8 (c_{12} C_2)^{1/2} n d_n^{-3/2}$, we obtain that for some positive constant $c_4$, 
 \begin{equation}
 P\{  ||({P}_n \{ \mathbf{B}_j(Y) \mathbf{B}_j^T(X_j) \})||  \geq (c_8 +1) (c_{12} C_2/d_n)^{1/2} \}  \leq 2 d_n^2 \exp ( -c_4 n d_n^{-2} )
 \label{eq:PYX}
\end{equation}

Next we deal with $|| \widehat{\mathbf{D}_j}^T ||$. Using Bernstein's inequality, we obtain that,
\begin{equation}
P\{|\widehat{b_{jm}} - b_{jm}| \geq \delta/n \} \leq 2 \exp \{- \frac{\delta^2}{2(C_2 n d_n^{-1} + 2\delta/3)}  \}
\label{eq:b}
\end{equation}

Since $ b_{jm} \geq C_3 k^{-1}$, by taking $\delta = C_3 w_1 n d_n ^{-1}$ for $w_1 \in (0,1)$, we have that there exists some positive constant $c_5$ such that
\begin{equation}
P\{ \widehat{b_{jm}} \leq C_3 (1-w_1) d_n^{-1}  \} \leq 2 \exp ( -c_5 n d_n^{-1} )
\label{eq:bhat}
\end{equation}

For $\mathbf{u} = (u_1, \ldots, u_{d_n-1})^T \in R^{d_n-1}$ with $\sum_{i=1}^{d_n-1} u_i^2 = 1$,
\begin{equation}
\begin{aligned}
\mathbf{u}^T \widehat{\mathbf{D}_j} \widehat{\mathbf{D}_j}^T \mathbf{u}  = \sum\limits_{m=1}^{d_n} \frac{1}{k^2 \widehat{b_{jm}}^2}  \left( \sum\limits_{i=1}^{d_n - 1} u_i z_{im} \right)^2 \leq \max\limits_{m} \frac{1}{k^2 \widehat{b_{jm}}^2}
\end{aligned}
\label{eq:Dhat}
\end{equation}

Combing (\ref{eq:b}), (\ref{eq:bhat}) and (\ref{eq:Dhat}), we have that
\begin{equation}
\begin{aligned}
P\{|| \widehat{\mathbf{D}_j}^T ||  \geq C_3^{-1} (1-w_1)^{-1} \} & \leq P\{\max\limits_{m} \frac{1}{k \widehat{b_{jm}}} \geq C_3^{-1} (1-w_1)^{-1} \}  \\
& \leq P\{\min\limits_{m} \widehat{b_{jm}} \leq C_3 (1-w_1) k^{-1} \}  \\
&\leq 2 d_n \exp ( -c_5 n d_n^{-1} )
\label{eq:Dh}
\end{aligned}
\end{equation}

Combining (\ref{eq:PYX}), (\ref{eq:Dh}), and $||\widehat{\mathbf{A}_{j0X}} || \leq ||{P}_n \{ \mathbf{B}_j(Y) \mathbf{B}_j^T(X_j) \}|| \hspace{0.05in} ||\widehat{\mathbf{D}_j}^T||$, we have
\begin{equation}
\begin{aligned}
& \qquad P\{||\widehat{\mathbf{A}_{j0X}} ||   \geq (c_8 +1)(c_{12} C_2)^{1/2} d_n^{-1/2} C_3^{-1} (1-w_1)^{-1} \} \\
& \leq P\{  ||{P}_n \{ \mathbf{B}_j(Y) \mathbf{B}_j^T(X_j) \}||  \geq (c_8 +1) (c_{12} C_2)^{1/2} d_n^{-1/2} \} + P\{|| \widehat{\mathbf{D}_j}^T ||  \geq C_3^{-1} (1-w_1)^{-1} \} \\
& \leq 2 d_n^2 \exp ( -c_4 n d_n^{-2} ) + 2 d_n \exp ( -c_5 n d_n^{-1} )
\label{eq:R4s}
\end{aligned}
\end{equation}

Result in {R4} follows by choosing $c_6$, $c_7$ accordingly.
\end{proof}

{R5}. \quad  There exist some positive constants $c_{9}$, $c_{10}$ such that, for any $\delta > 0$, 
\begin{equation}
\begin{aligned}
P\{||\widehat{\mathbf{A}_{j0X}} & - \mathbf{A}_{j0X}|| \geq c_{9} d_n^2 \delta^2/n^2  + c_{10} d_n \delta/n \} \\ &\leq 8 d_n^2 \exp \{- \frac{\delta^2}{2(C_2 n d_n^{-1} + 2\delta/3)}  \} +  4 d_n \exp ( -c_5 n d_n^{-1} )
\end{aligned}
 \label{eq:R5}
\end{equation}

\begin{proof} \quad It is easy to derive
\begin{equation}
\begin{aligned}
& ||\widehat{\mathbf{A}_{j0X}} - \mathbf{A}_{j0X}|| = || {P}_n \{ \mathbf{B}_j(Y) \mathbf{B}_j^T(X_j) \} \widehat{\mathbf{D}_j}^T -  {E}\{ \mathbf{B}_j(Y) \mathbf{B}_j^T(X_j) \} {\mathbf{D}_j}^T ||\\
& \leq ||({P}_n -{E}) \{\mathbf{B}_j(Y) \mathbf{B}_j^T(X_j) \}|| \hspace{0.05in} || \widehat{\mathbf{D}_j}^T - \mathbf{D}_j^T|| +  ||{E}\{ \mathbf{B}_j(Y) \mathbf{B}_j^T(X_j) \}|| \hspace{0.05in} || \widehat{\mathbf{D}_j}^T - \mathbf{D}_j^T||\\
& \qquad \qquad + ||({P}_n -{E}) \{\mathbf{B}_j(Y) \mathbf{B}_j^T(X_j) \}|| \hspace{0.05in} || {\mathbf{D}_j}^T ||
\end{aligned}
\label{eq:R5all}
\end{equation}

It is proved in {R2} that $|| {E} \{ \mathbf{B}_{j}(Y) \mathbf{B}_j^T(X_j) \}|| \leq (c_{12} C_2/ d_n)^{1/2}$ and that $|| {\mathbf{D}_j}^T || \leq C_3^{-1}$. Combining (\ref{eq:R4a}) and the fact that
\begin{equation*}
||({P}_n - {E}) \{ \mathbf{B}_j(Y) \mathbf{B}_j^T(X_j) \}|| \leq d_n ||({P}_n - {E}) \{ \mathbf{B}_j(Y) \mathbf{B}_j^T(X_j) \}||_{sup},
\end{equation*}   we have that,  
\begin{equation}
P\{||({P}_n -{E}) \{\mathbf{B}_j(Y) \mathbf{B}_j^T(X_j) \}|| \geq d_n \delta/n \} \leq 2 d_n^2 \exp \{- \frac{\delta^2}{2(C_2 n d_n^{-1} + 2\delta/3)} \}.
\label{eq:R5b}
\end{equation}

For $\mathbf{u} \in R^{d_n-1}$ with $\sum_{i=1}^{d_n-1} u_i^2 = 1$,
\begin{equation}
\begin{aligned}
\mathbf{u}^T (\widehat{\mathbf{D}_j} - \mathbf{D}_j) (\widehat{\mathbf{D}_j} - \mathbf{D}_j) ^T \mathbf{u} & = \sum\limits_{m=1}^{d_n} \left( \frac{1}{k \widehat{b_{jm}}} - \frac{1}{k b_{jm}} \right)^2 \left( \sum\limits_{i=1}^{d_n - 1} u_i z_{im} \right)^2\\
& \leq C_3^{-2} \mathop{\max}\limits_{m} \frac{(\widehat{b_{jm}} - b_{jm})^2}{\widehat{b_{jm}}^2}
\end{aligned}
\label{eq:DD}
\end{equation}

From (\ref{eq:b}), (\ref{eq:bhat}) and (\ref{eq:DD}), we have that,
\begin{equation}
\begin{aligned}
& P\{|| \widehat{\mathbf{D}_j}^T - \mathbf{D}_j^T||  \geq C_3^{-2}  (1-w_1)^{-1} d_n \delta/n \} \\
& \leq P\{C_3^{-1} \mathop{\max}\limits_{m} \frac{|\widehat{b_{jm}} - b_{jm}|}{\widehat{b_{jm}}}
  \geq C_3^{-1} \frac{\delta/n}{ C_3 (1-w_1) d_n^{-1}} \} \\
& \leq  P\{ \mathop{\max}\limits_{m} |\widehat{b_{jm}} - b_{jm}| \geq \delta/n \}  +  P\{\mathop{\min}\limits_{m} \widehat{b_{jm}} \leq  C_3 (1-w_1) d_n^{-1} \} \\
& \leq 2 d_n \exp \{- \frac{\delta^2}{2(C_2 n d_n^{-1} + 2\delta/3)}  \} + 2 d_n \exp ( -c_5 n d_n^{-1} ) 
\label{eq:D}
\end{aligned}
\end{equation}

Therefore, together with (\ref{eq:R5all}), (\ref{eq:R5b}), (\ref{eq:D}) and union bound of probability, we have
\begin{equation*}
\begin{aligned}
& P\{  ||\widehat{\mathbf{A}_{j0X}} - \mathbf{A}_{j0X}|| \geq \frac{d_n^2 \delta^2/n^2}{C_3^2 (1-w_1) }  + \frac{(c_{12} C_2)^{1/2} d_n^{1/2} \delta/n}{C_3^2 (1-w_1) } + C_3^{-1} d_n \delta/n \} \\
& \leq 4 d_n^2 \exp \{- \frac{\delta^2}{2(C_2 n d_n^{-1} + 2\delta/3)}  \}  + 4 d_n \exp \{- \frac{\delta^2}{2(C_2 n d_n^{-1} + 2\delta/3)}  \}  + 4 d_n \exp ( -c_5 n d_n^{-1} )
\end{aligned}
\end{equation*}
  
Result in {R5} can be obtained by adjusting the values of $c_{9}$ and $c_{10}$.
\end{proof}

{R6}. \quad For given $c_4$ and $c_5$, there exist positive constants $c_{15}$ and $ c_{16}$ such that,
\begin{equation}
\begin{aligned}
  P\{ ||& \widehat{\mathbf{A}_{jXX}}^{-1}|| \geq c_{16}  d_n \} \\ & \leq 2 d_n^2 \exp ( -c_4 n d_n^{-3} ) + 2 d_n^3 \exp (-c_{15} n d_n^{-7} ) + 2 d_n^3 \exp ( -c_5 n d_n^{-1} )
\end{aligned}
\label{eq:R6}
\end{equation}

\begin{proof} \quad Follow the proof in Lemma 5  of \cite{FanFengSong11NIS}, we have that,
\begin{equation*}
|\lambda_{min}(\widehat{\mathbf{D}_j} \widehat{\mathbf{D}_j}^T) - \lambda_{min}(\mathbf{D}_j \mathbf{D}_j^T) |  \leq d_n ||\mathbf{V}_j||_{sup}, \mbox{ where } \mathbf{V}_j = \widehat{\mathbf{D}_j} \widehat{\mathbf{D}_j}^T - \mathbf{D}_j \mathbf{D}_j^T 
\end{equation*}

The $(s,m)$-entry of $\mathbf{V}_j$ is 
\begin{equation*}
\begin{aligned}
\left(\mathbf{V}_j \right) ^{(s,m)} &= |\sum\limits_{i=1}^{d_n} \frac{z_{si} z_{mi}}{k^2} \left( \frac{1}{ \widehat{b_{ji}}^2} - \frac{1}{b_{ji}^2} \right)| = |\sum\limits_{i=1}^{d_n} \frac{z_{si} z_{mi}}{k^2 b_{ji}^2} \left( \frac{b_{ji}^2 - \widehat{b_{ji}}^2}{ \widehat{b_{ji}}^2} \right)|\\
 &\leq C_3^{-2} d_n \max\limits_{i} | \frac{b_{ji}^2 - \widehat{b_{ji}}^2}{ \widehat{b_{ji}}^2}| \leq 2 C_3^{-2} d_n \max\limits_{i} | \frac{ b_{ji} - \widehat{b_{ji}}}{ \widehat{b_{ji}}^2}|
\end{aligned}
\end{equation*}

It is clear that $||\mathbf{V}_j||_{sup} \leq 2 C_3^{-2} d_n \max\limits_{i} | ( b_{ji} - \widehat{b_{ji}})/{ \widehat{b_{ji}}^2}|$. 
Together with (\ref{eq:b}) and (\ref{eq:bhat}) , we have
\begin{equation*}
\begin{aligned}
& P\{ |\lambda_{min}(\widehat{\mathbf{D}_j} \widehat{\mathbf{D}_j}^T)  - \lambda_{min}(\mathbf{D}_j \mathbf{D}_j^T) |  \geq 2 C_3^{-4}(1-w_1)^{-2} d_n^4 \delta/n   \} \\
& \leq P\{  2 C_3^{-2} d_n^2 \max\limits_{i} | \frac{ b_{ji} - \widehat{b_{ji}}}{ \widehat{b_{ji}}^2}|  \geq 2 C_3^{-2} d_n^2 \delta/n \times {C_3^{-2} (1-w_1)^{-2} d_n^{2}}  \} \\
& \leq  P\{ \mathop{\max}\limits_{m} |\widehat{b_{jm}} - b_{jm}| \geq \delta/n \}  +  P\{\mathop{\min}\limits_{m} \widehat{b_{jm}} \leq  C_3 (1-w_1) d_n^{-1} \} \\
& \leq 2 d_n \exp \{- \frac{\delta^2}{2(C_2 n d_n^{-1} + 2\delta/3)} \} + 2 d_n \exp ( -c_5 n d_n^{-1} )
\end{aligned}
\end{equation*}
which indicates that there exists a positive constant $c_{14}$,
\begin{equation}
\begin{aligned}
P\{ |\lambda_{min}(\widehat{\mathbf{D}_j} \widehat{\mathbf{D}_j}^T) - &\lambda_{min}(\mathbf{D}_j \mathbf{D}_j^T) |  \geq c_{14} d_n^4 \delta/n   \} \\& \leq 2 d_n \exp \{- \frac{\delta^2}{2(C_2 n d_n^{-1} + 2\delta/3)} \} + 2 d_n \exp (-c_5 n d_n^{-1} )
\label{eq:DDhat}
\end{aligned}
\end{equation}

Due to the facts that
\begin{equation*}
\begin{aligned}
  c_{11} k^{-1} \leq \lambda_{min}(\mathbf{D}_j {E} \{& \mathbf{B}_j(X_j) \mathbf{B}_j^T(X_j) \} \mathbf{D}_j^T ) \leq \\& \lambda_{max}({E} \{ \mathbf{B}_j(X_j) \mathbf{B}_j^T(X_j) \}) \lambda_{min}(\mathbf{D}_j \mathbf{D}_j^T) \leq c_{12} k^{-1} \lambda_{min}(\mathbf{D}_j \mathbf{D}_j^T)
\end{aligned}
\end{equation*}
and that
\begin{equation*}
\begin{aligned}
  c_{11} k^{-1} \lambda_{max}(\mathbf{D}_j \mathbf{D}_j^T) \leq \lambda_{min}({E} \{ \mathbf{B}_j(X_j)& \mathbf{B}_j^T(X_j) \}) \lambda_{max}(\mathbf{D}_j \mathbf{D}_j^T) \leq \\ & \lambda_{max}(\mathbf{D}_j {E} \{ \mathbf{B}_j(X_j) \mathbf{B}_j^T(X_j) \} \mathbf{D}_j^T ) \leq c_{12} k^{-1}
\end{aligned}
\end{equation*}
we have
\begin{equation*}
\begin{aligned}
 \frac{c_{11}}{c_{12}} \leq  \lambda_{min}(\mathbf{D}_j \mathbf{D}_j^T) \leq \lambda_{max}(\mathbf{D}_j \mathbf{D}_j^T) \leq  \frac{c_{12}}{c_{11}}
\end{aligned}
\end{equation*}

By taking $\delta = w_2/c_{14} n d_n^{-4} \times c_{11}/c_{12}$ in (\ref{eq:DDhat}) for any $w_2 \in (0,1)$,  there exists a positive constant $c_{15}$ such that, 
\begin{equation*}
\begin{aligned}
P\{ |\lambda_{min}(\widehat{\mathbf{D}_j} \widehat{\mathbf{D}_j}^T) - \lambda_{min}(\mathbf{D}_j& \mathbf{D}_j^T) |  \geq w_2 \lambda_{min}(\mathbf{D}_j \mathbf{D}_j^T) \} \\ & \leq 2 d_n \exp (-c_{15} n d_n^{-7} ) + 2 d_n \exp ( -c_5 n d_n^{-1} )
\end{aligned}
\end{equation*}

By following a similar argument in proving inequality (26) in NIS \citep{FanFengSong11NIS}, we have,
\begin{equation}
 P\{ \lambda_{min}^{-1}(\widehat{\mathbf{D}_j} \widehat{\mathbf{D}_j}^T) \geq (c_8 +1)c_{12}/c_{11} \}  \leq 2 d_n \exp (-c_{15} n d_n^{-7} ) + 2 d_n \exp ( -c_5 n d_n^{-1} )
\label{eq:Dhatmin}
\end{equation}

Similarly, it is easy to obtain 
\begin{equation}
\begin{aligned}
 P\{ \lambda_{min}^{-1}({P}_n \{ \mathbf{B}_j(X_j) \mathbf{B}_j^T(X_j) \}) \geq (c_8 +1)c_{11}^{-1} d_n \}  \leq 2 d_n^2 \exp ( -c_4 n d_n^{-3} )
\label{eq:PXXinv}
\end{aligned}
\end{equation}

Due to the fact that $\lambda_{max}({\mathbf{H}}^{-1}) = \lambda_{min}^{-1}({\mathbf{H}})$, we have
\begin{equation*}
||\widehat{\mathbf{A}_{jXX}}^{-1}|| = \lambda_{min}^{-1}(\widehat{\mathbf{A}_{jXX}}) \leq \lambda_{min}^{-1}({P}_n \{ \mathbf{B}_j(X_j) \mathbf{B}_j^T(X_j) \}) \hspace{0.05in} \lambda_{min}^{-1}(\widehat{\mathbf{D}_j} \widehat{\mathbf{D}_j}^T) 
\end{equation*}  

Together with  (\ref{eq:Dhatmin}) and (\ref{eq:PXXinv}), we can obtain that
\begin{equation*}
\begin{aligned}
& P\{ || \widehat{\mathbf{A}_{jXX}}^{-1}|| \geq (c_8 +1)^2 c_{12} c_{11}^{-2} d_n \} \\
& \leq P\{ \lambda_{min}^{-1}({P}_n \{ \mathbf{B}_j(X_j) \mathbf{B}_j^T(X_j) \}) \hspace{0.05in} \lambda_{min}^{-1}(\widehat{\mathbf{D}_j} \widehat{\mathbf{D}_j}^T)  \geq (c_8 +1)^2 c_{12} c_{11}^{-2} d_n \} \\
& \leq P\{ \lambda_{min}^{-1}({P}_n \{ \mathbf{B}_j(X_j) \mathbf{B}_j^T(X_j) \}) \geq (c_8 +1)c_{12}/c_{11} \} +  P\{ \lambda_{min}^{-1}(\widehat{\mathbf{D}_j} \widehat{\mathbf{D}_j}^T)  \geq (c_8 +1)c_{11}^{-1} d_n \} \\
& \leq 2 d_n^2 \exp ( -c_4 n d_n^{-3} ) + 2 d_n \exp (-c_{15} n d_n^{-7} ) + 2 d_n \exp ( -c_5 n d_n^{-1} )
\end{aligned}
\end{equation*}

Therefore, {R6} follows by choosing $c_{16} = (c_8 +1)^2 c_{12} c_{11}^{-2}$. \end{proof}

{R7}. \quad For any $\delta > 0$, given positive constant $c_4$, there exists a positive constant $c_{17}$ such that,
\begin{equation}
 P\{ || \widehat{\mathbf{A}_{j00}}^{-1/2} - \mathbf{A}_{j00}^{-1/2}||
 \geq c_{17} d_n^{5/2} \delta/n \}  \leq 2 d_n^2 \exp ( -c_{4} n d_n^{-3} ) +  2 d_n^2 \exp \{- \frac{\delta^2}{2(C_2 n d_n^{-1} + 2\delta/3)} \}
 \label{eq:R7} 
\end{equation} 

\begin{proof} 
\quad  Using perturbation theory \citep{kato1995}, it is proved \citep*[\textit{Lemma 6.3}]{Burman91ACE} that for some $c_{18} > 0$,
\begin{equation}
|| \widehat{\mathbf{A}_{j00}}^{-1/2} - \mathbf{A}_{j00}^{-1/2}|| \leq c_{18} \tilde{\gamma}^{-3/2} ||\widehat{\mathbf{A}_{j00}} - \mathbf{A}_{j00}||
\label{eq:R7a}
\end{equation}
where $\tilde{\gamma}$ is the minimum of the smallest eigenvalues of $\widehat{\mathbf{A}_{j00}}$ and $\mathbf{A}_{j00}$. $\tilde{\gamma}$ is positive by definition.  Therefore, 
\begin{equation*}
\tilde{\gamma}^{-1} = \max \{ \lambda_{min}^{-1} (\widehat{\mathbf{A}_{j00}} ), \lambda_{min}^{-1} (\mathbf{A}_{j00} ) \} = \max \{ ||\widehat{\mathbf{A}_{j00}}^{-1}||, ||\mathbf{A}_{j00}^{-1}|| \}
\end{equation*}

From \textit{Fact 3} and {R3}, we have,
\begin{subequations}
\begin{align} 
& c_{12}^{-1} d_n \leq ||\mathbf{A}_{j00}^{-1}|| \leq c_{11}^{-1} d_n
\label{eq:R71}\\
& P\{ ||[{P}_n \{\mathbf{B}_j(Y) \mathbf{B}_j^T(Y) \}]^{-1}||  \geq (c_8 +1) c_{11}^{-1} d_n \}  \leq 2 d_n^2 \exp ( -c_4 n d_n^{-3} )
\label{eq:R72}
\end{align} 
\end{subequations}

Combining (\ref{eq:R71}) and (\ref{eq:R72}) yields
\begin{equation*}
 P\{ \tilde{\gamma}^{-1} \geq \max \left( (c_8 +1) c_{11}^{-1} d_n, c_{11}^{-1} d_n \right) \} \leq 2 d_n^2 \exp (-c_{4} n d_n^{-3} ) 
\end{equation*}
which is,
\begin{equation}
 P \{ \tilde{\gamma}^{-1} \geq  (c_8 +1) c_{11}^{-1} d_n \} \leq 2 d_n^2 \exp ( -c_{4} n d_n^{-3} )
\label{eq:R7r}
\end{equation}

Additionally, as proved in equation (33) in \cite{FanFengSong11NIS}, we have large deviation bound for $||({P}_n - {E}) \{ \mathbf{B}_j(Y)\mathbf{B}_j^T(Y) \}||$,
\begin{equation}
\begin{aligned}
P\{||({P}_n - {E}) \{ \mathbf{B}_j(Y)\mathbf{B}_j^T(Y) \}|| \geq  d_n \delta/n \} \leq 2 d_n^2 \exp \{- \frac{\delta^2}{2(C_2 n d_n^{-1} + 2\delta/3)}  \} 
\end{aligned}
\label{eq:33Y}
\end{equation} 
 
By (\ref{eq:R7a}), (\ref{eq:R7r}), (\ref{eq:33Y}) and under the union bound of probability, we have that,

\begin{equation}
\begin{aligned}
& P\{ || \widehat{\mathbf{A}_{j00}}^{-1/2} - \mathbf{A}_{j00}^{-1/2}|| \geq c_{18} (c_8 +1)^{3/2} c_{11}^{-3/2} d_n^{5/2}\delta/n \} \\ 
& \leq P\{ c_{18} \tilde{\gamma}^{-3/2} ||\widehat{\mathbf{A}_{j00}} - \mathbf{A}_{j00}|| \geq c_{18} (c_8 +1)^{3/2} c_{11}^{-3/2} d_n^{3/2} \hspace{0.05in} d_n\delta/n \} \\
&\leq P\{ \tilde{\gamma}^{-1} \geq (c_8 +1) c_{11}^{-1} d_n \}  + P\{  || \widehat{ \mathbf{A}_{j00}} - \mathbf{A}_{j00}|| \geq d_n\delta/n \} \\
& \leq 2 d_n^2 \exp ( -c_{4} n d_n^{-3} ) +  2 d_n^2 \exp \{- \frac{\delta^2}{2(C_2 n d_n^{-1} + 2\delta/3)} \}
\end{aligned}
\end{equation} 

Therefore, {R7} follows by choosing $c_{17} = c_{18} (c_8 +1)^{3/2} c_{11}^{-3/2}$.
\end{proof}

{R8}. \quad For any $\delta > 0$, given positive constant $c_4$, there exist a positive constant $c_{19}$ such that,
\begin{equation}
\begin{aligned}
  P\{ || \widehat{\mathbf{A}_{jXX}}^{-1} - \mathbf{A}_{jXX}^{-1}||
 & \geq c_{19} (d_n^5 \delta^3/n^3 + d_n^3 \delta/n) \}  \leq  8 d_n^2  \exp \{- \frac{\delta^2}{2(C_2 n d_n^{-1} + 2\delta/3)} \} + \\ & 4 d_n^2 \exp ( -c_{4} n d_n^{-3} ) +2 d_n \exp (-c_{15} n d_n^{-7} ) + 6 d_n \exp ( -c_5 n d_n^{-1} ) 
 \label{eq:R8} 
\end{aligned}
\end{equation} 

\begin{proof} 
\quad It's obvious that
\begin{equation}
\begin{aligned}
||\widehat{\mathbf{A}_{jXX}}^{-1} - \mathbf{A}_{jXX}^{-1}|| \leq || \mathbf{A}_{jXX}^{-1}|| \hspace{0.05in} ||\mathbf{A}_{jXX} - \widehat{\mathbf{A}_{jXX}} || \hspace{0.05in} ||\widehat{\mathbf{A}_{jXX}}^{-1}||
\end{aligned}
\label{eq:R8de}
\end{equation}
and that 
\begin{equation}
\begin{aligned}
&||\widehat{\mathbf{A}_{jXX}} - \mathbf{A}_{jXX}|| = || \widehat{\mathbf{D}_j} {P}_n \{ \mathbf{B}_j(X_j) \mathbf{B}_j^T(X_j) \} \widehat{\mathbf{D}_j}^T -  \mathbf{D}_j {E} \{ \mathbf{B}_j(X_j) \mathbf{B}_j^T(X_j) \} \mathbf{D}_j^T ||\\
& \leq ||\widehat{\mathbf{D}_j} - \mathbf{D}_j|| \hspace{0.05in} ||({P}_n -{E})  \{ \mathbf{B}_j(X_j) \mathbf{B}_j^T(X_j) \}|| \hspace{0.05in} || \widehat{\mathbf{D}_j}^T - \mathbf{D}_j^T|| + 2 ||{P}_n \{ \mathbf{B}_j(X_j) \mathbf{B}_j^T(X_j) \}|| \times \\
& \qquad   || \widehat{\mathbf{D}_j}^T - \mathbf{D}_j^T|| + || \mathbf{D}_j^T || \hspace{0.05in} ||({P}_n -{E})  \{ \mathbf{B}_j(X_j) \mathbf{B}_j^T(X_j) \}|| \hspace{0.05in} || \mathbf{D}_j ||
\end{aligned}
\label{eq:R8all}
\end{equation}

From the similar reasoning in proving (\ref{eq:PYX}) and (\ref{eq:R5b}), it is easy to obtain that
\begin{equation}
 P\{  ||{P}_n \{ \mathbf{B}_j(X_j) \mathbf{B}_j^T(X_j) \}||  \geq (c_8 +1)c_{13} d_n^{-1} \}  \leq 2 d_n^2 \exp ( -c_4 n d_n^{-3} )
 \label{eq:PXX}
\end{equation}
\begin{equation}
P\left(||({P}_n -{E})  \{ \mathbf{B}_j(X_j) \mathbf{B}_j^T(X_j) \}|| \geq d_n \delta/n \right) \leq 2 d_n^2 \exp \{- \frac{\delta^2}{2(C_2 n d_n^{-1} + 2\delta/3)}  \}
\label{eq:R8a}
\end{equation} 

With $c_{19}$ chosen properly, results in {R8} follows by combining \textit{Fact 3}, (\ref{eq:D}), (\ref{eq:R6}), (\ref{eq:R8de}), (\ref{eq:R8all}), (\ref{eq:PXX}), (\ref{eq:R8a}) and the fact $||\mathbf{D}_j^T|| < C_3^{-1}$.
\end{proof}

\subsection{Proof of Theorem 1}
\textit{Proof of Theorem 1.} 
Recall that 

\begin{equation*}
\lambda_{j1}^* = || \mathbf{A}_{j00}^{-1/2} \mathbf{A}_{j0X} \mathbf{A}_{jXX}^{-1}  \mathbf{A}_{jX0} \mathbf{A}_{j00}^{-1/2}||
\end{equation*}
and that
\begin{equation*}
\widehat{\lambda_{j1}^*} = || \widehat{\mathbf{A}_{j00}}^{-1/2} \widehat{\mathbf{A}_{j0X}} \widehat{\mathbf{A}_{jXX}}^{-1}  \widehat{\mathbf{A}_{j0X}}^T \widehat{\mathbf{A}_{j00}}^{-1/2}||
\end{equation*}

Let 
$\mathbf{a} = \mathbf{A}_{j00}^{-1/2}$, 
$\mathbf{b} = \mathbf{A}_{j0X}$, 
$\mathbf{H}=  \mathbf{A}_{jXX}^{-1}$,  
$\mathbf{a_n} = \widehat{\mathbf{A}_{j00}}^{-1/2}$,
$\mathbf{b_n} = \widehat{\mathbf{A}_{j0X}} $,
$\mathbf{H_n} =\widehat{\mathbf{A}_{jXX}}^{-1}$,

\begin{equation}
\begin{aligned}
& \widehat{\lambda_{j1}^*} - \lambda_{j1}^* = ||\mathbf{a_n}^T \mathbf{b_n}^T \mathbf{H_n} \mathbf{b_n} \mathbf{a_n}|| - ||\mathbf{a}^T \mathbf{b}^T \mathbf{H} \mathbf{b} \mathbf{a}||\\
& \leq ||(\mathbf{a}_n - \mathbf{a})^T \mathbf{b}_n^T \mathbf{H}_n \mathbf{b}_n (\mathbf{a}_n - \mathbf{a})|| + 2 || (\mathbf{a}_n - \mathbf{a})^T \mathbf{b}_n^T \mathbf{H}_n \mathbf{b}_n \mathbf{a} || + ||\mathbf{a}^T (\mathbf{b}_n^T \mathbf{H}_n \mathbf{b}_n - \mathbf{b}^T \mathbf{H} \mathbf{b}) \mathbf{a}|| \\
& \qquad  \triangleq S_1 + S_2 + S_3 
\end{aligned}
\label{eq:S}
\end{equation}

We denote the terms in r.h.s as $S_1$, $S_2$ and $S_3$ respectively.  Furthermore, we let the r.h.s of inequalities (\ref{eq:R4}),(\ref{eq:R5}),(\ref{eq:R6}),(\ref{eq:R7}),(\ref{eq:R8}) as $Q_4$, $Q_5$, $Q_6$, $Q_7$, $Q_8$.\\ 

Note that 
\begin{equation}
S_1 \leq ||\mathbf{a}_n - \mathbf{a}||^2 \hspace{0.05in} ||\mathbf{b}_n||^2 \hspace{0.05in} ||\mathbf{H}_n|| 
\label{eq:S1}
\end{equation}

By (\ref{eq:R4}),(\ref{eq:R6}),(\ref{eq:R7}), we have that there exist a positive constant $c_{20}$ such that, 
\begin{equation}
\begin{aligned}
P\{S_1 \geq c_{20} d_n^5 \delta^2/n^2\} \leq Q_4 + Q_6 + Q_7
\end{aligned}
\label{eq:S1S}
\end{equation}

As to $S_2$,
\begin{equation}
S_2 \leq ||\mathbf{a}_n - \mathbf{a}|| \hspace{0.05in} ||\mathbf{b}_n||^2 \hspace{0.05in} ||\mathbf{H}_n|| \hspace{0.05in} ||\mathbf{a}|| 
\label{eq:S2}
\end{equation}

By (\ref{eq:R1}),(\ref{eq:R4}),(\ref{eq:R6}),(\ref{eq:R7}), we have that there exist a positive constant $c_{21}$ such that, 
\begin{equation}
\begin{aligned}
P\{S_2 \geq c_{21} d_n^3 \delta/n \} \leq Q_4 + Q_6 + Q_7
\label{eq:S2S}
\end{aligned}
\end{equation}

As to $S_3$,
\begin{equation}
\begin{aligned}
S_3 &\leq ||\mathbf{a}||^2 \hspace{0.05in} ||\mathbf{b}_n^T \mathbf{H}_n B_n - \mathbf{b}^T \mathbf{H} \mathbf{b}||\\
& \leq ||\mathbf{a}||^2 (||(\mathbf{b}_n - \mathbf{b})^T  \mathbf{H}_n  (\mathbf{b}_n - \mathbf{b})|| + 2 || (\mathbf{b}_n - \mathbf{b})^T \mathbf{H}_n \mathbf{b} || + ||\mathbf{b}^T (\mathbf{H}_n - \mathbf{H}) \mathbf{b}||)\\
& \qquad  \triangleq  ||\mathbf{a}||^2 (S_{31} + 2S_{32} + S_{33})
\end{aligned}
\label{eq:S3}
\end{equation}

Note that 
\begin{equation}
S_{31} \leq ||\mathbf{b}_n - \mathbf{b}||^2 \hspace{0.05in} ||\mathbf{H}_n|| 
\label{eq:S31}
\end{equation}

By (\ref{eq:R5}),(\ref{eq:R6}), we have that there exist a positive constant $c_{22}$ such that, 
\begin{equation}
\begin{aligned}
P\{S_{31} \geq c_{22} d_n^5 (\delta^2/n^2 +\delta/n)^2\} \leq Q_5 + Q_6
\end{aligned}
\label{eq:S31S}
\end{equation}

As to $S_{32}$,
\begin{equation}
S_{32} \leq ||\mathbf{b}_n - \mathbf{b}|| \hspace{0.05in} ||\mathbf{H}_n|| \hspace{0.05in} ||\mathbf{b}|| 
\label{eq:S32}
\end{equation}

By (\ref{eq:R2}),(\ref{eq:R5}),(\ref{eq:R6}),(\ref{eq:R7}), we have that there exist a positive constant $c_{23}$ such that, 
\begin{equation}
\begin{aligned}
P\{S_{32} \geq c_{23} d_n^{5/2} (\delta^2/n^2 +\delta/n) \} \leq Q_5 + Q_6
\label{eq:S32S}
\end{aligned}
\end{equation}

As to $S_{33}$,
\begin{equation}
S_{33} \leq ||\mathbf{b}||^2 \hspace{0.05in} ||\mathbf{H}_n - \mathbf{H}|| 
\label{eq:S33}
\end{equation}

By (\ref{eq:R2}),(\ref{eq:R8}), we have that there exist a positive constant $c_{24}$ such that, 
\begin{equation}
\begin{aligned}
P\{S_{33} \geq c_{24}(d_n^4 \delta^3/n^3 + d_n^2 \delta/n) \} \leq Q_8 
\label{eq:S33S}
\end{aligned}
\end{equation}

Combining (\ref{eq:R1}),(\ref{eq:S3}),(\ref{eq:S31}),(\ref{eq:S32}),(\ref{eq:S33}), we have 
\begin{equation}
\begin{aligned}
P\{S_3 \geq c_{22} & d_n^6 (\delta^2/n^2 +\delta/n)^2 + c_{23} d_n^{7/2}(\delta^2/n^2 +\delta/n)  +  c_{24}(d_n^5 \delta^3/n^3 + d_n^3 \delta/n)\} \\
& \leq 2 Q_5 + 2 Q_6 + Q_8 
\label{eq:S3S}
\end{aligned}
\end{equation}

Define $\varsigma(d_n, \delta) =  c_{20} d_n^5 \delta^2/n^2 + c_{21} d_n^3 \delta/n + c_{22} d_n^6 (\delta^2/n^2 +\delta/n)^2 + c_{23}  d_n^{7/2} (\delta^2/n^2 +\delta/n)  +  c_{24}(d_n^5 \delta^3/n^3 +  d_n^{3} \delta/n)$. Then from (\ref{eq:S}),(\ref{eq:S1S}),(\ref{eq:S2S}),(\ref{eq:S3S}), we have that due to symmetry, %
\begin{equation}
\begin{aligned}
P\{ |\widehat{\lambda_{j1}^*} - \lambda_{j1}^*| \geq \varsigma(d_n, \delta) \} \leq 4 Q_4 +  4 Q_5 + 8 Q_6 + 4 Q_7 + 2 Q_8 
\label{eq:SS}
\end{aligned}
\end{equation}

By properly choosing the value of $\delta$ (i.e., taking $\delta = {c_2}{(c_{22} +c_{23})^{-1}} d_n^{-5/2} n ^{1-2\kappa}$), we can make $\varsigma(d_n, \delta) = c_2 d_n n^{-2\kappa}$, for any $c_2>0$.  Then, we have
\begin{equation}
\begin{aligned}
P( |\widehat{\lambda_{j1}^*} - \lambda_{j1}^*| \geq c_2 d_n n^{-2\kappa} ) \leq \mathcal{O} \left( d_n^2 \exp ( - c_3 n^{1-4\kappa} d_n^{-4} )  +  d_n \exp ( -c_4 n d_n^{-7} ) \right) 
\label{eq:SS1}
\end{aligned}
\end{equation}

The first part of Theorem 1 follows via the union bound of probability.

To prove the second part, we define a event 
\begin{equation*}
 \mathcal{A}_n \equiv \lbrace \mathop{max}\limits_{j \in \mathcal{D}} 
|\widehat{\lambda_{j1}^*} - \lambda_{j1}^*|  
\leq c_1 \xi d_n n^{-2\kappa} /2 \rbrace
\end{equation*}

By Lemma \ref{lemma1}, we have  
\begin{equation}
 \widehat{\lambda_{j1}^*} \geq c_1 \xi d_n n^{-2\kappa} /2 , \forall j \in \mathcal{D}
\label{eq:36}
\end{equation}

Thus, by choosing $\nu_n = c_5 d_n n^{-2\kappa}$ with $c_5 \leq c_1 \xi/2$. We have that $\mathcal{D} \subseteq \widehat{\mathcal{D}_{\nu_n}}$.  Therefore,
\begin{equation*}
P (\mathcal{A}_n^c)  \leq \mathcal{O} \left( s \{ d_n^2 \exp \lbrace - c_3 n^{1-4\kappa} d_n^{-4} \rbrace  +  d_n \exp ( -c_4 n d_n^{-7} ) \} \right) 
\end{equation*}

Then the probability bound for the second part of Theorem 1 is attained. 
{
\subsection{Proof sketch of Theorem 2}
\textit{Proof of Theorem 2.} 
From subsection \ref{subsec:ot}, we have that $ \lambda_{j1}^* =  {E} (\phi^{*2}_{nj}) $ and $ \widehat{\lambda_{j1}^*} =  {P_n} (\phi^{*2}_{nj}) $.

From equation (\ref{eq:opt_bs0}), after obtaining $\theta^*_{nj}$ where $\mbox{Var}(\theta^{*}_{nj}) = 1$, $\phi^*_{nj}$ can be obtained via the following optimization problem.
\begin{equation*}
\begin{aligned}
& \underset{ \phi_{nj} \in \mathcal{S}_n}{\text{arg min}}
& &  {E}[\{ \theta^*_{nj}(Y) - \phi_{nj}(X_j)\}^2], \mbox{ where $\phi_{nj}(X_j) =  {\boldsymbol\eta}_j^T \boldsymbol{\psi}_j(X_j)$.}
\end{aligned}
\end{equation*}

Therefore, $\phi^*_{nj} = \boldsymbol{\psi}_j^T  E \{ {\boldsymbol\psi_j} {\boldsymbol\psi_j}^T\} ^{-1} E\boldsymbol{\psi}_j \theta^{*}_{nj} $.

We notice that the only difference between our proof and the proof of Theorem 2 in \cite{FanFengSong11NIS} is the role of $Y$.  As MC-SIS essentially uses transformation of $Y$, we can not deal directly with $Y$.  However, from the formulation above, $\theta_{nj}^*$ here plays the same role as $Y$ in \cite{FanFengSong11NIS}.  With this connection,  our proof follows immediately by replacing $Y$ in the proof of Theorem 2 in \cite{FanFengSong11NIS} with  $\theta_{nj}^*$.
}

\bibliographystyle{apalike}
\bibliography{paper}
\end{document}